\documentclass[aps,pra,preprint,superscriptaddress]{revtex4-2}

\usepackage{graphicx}% Include figure files
\usepackage{latexsym}
\usepackage{dcolumn}% Align table columns on decimal point
\usepackage{bm}% bold math
\usepackage[utf8]{inputenc}
\usepackage[T1]{fontenc}
\usepackage{tabularx}
\usepackage{hyperref}
\usepackage{color}
\usepackage{placeins}
\usepackage{float}
\usepackage{amsmath}
\usepackage{gensymb}
\usepackage{diagbox}

\begin{document}

\title{Stochastic dynamics simulation of the focused electron beam induced deposition process}

\author{Ilia A. Solov'yov}
\affiliation{Institute of Physics, Carl von Ossietzky Universit{\"a}t Oldenburg, Carl-von-Ossietzky-Str. 9-11, 26129 Oldenburg Germany}
\affiliation{Research Centre for Neurosensory Science, Carl von Ossietzky Universit{\"a}t Oldenburg, Carl-von-Ossietzky-Str. 9-11, 26129 Oldenburg, Germany}
\affiliation{Center for Nanoscale Dynamics (CENAD), Carl von Ossietzky Universit{\"a}t Oldenburg, Ammerl{\"a}nder Heerstr. 114-118, 26129 Oldenburg, Germany}

\author{Alexey Prosvetov}
\affiliation{Institute of Physics, Carl von Ossietzky Universit{\"a}t Oldenburg, Carl-von-Ossietzky-Str. 9-11, 26129 Oldenburg Germany}
\affiliation{MBN Research Center, Altenh\"oferallee 3, 60438 Frankfurt am Main, Germany}

\author{Gennady Sushko}
\affiliation{MBN Research Center, Altenh\"oferallee 3, 60438 Frankfurt am Main, Germany}

\author{Andrey V. Solov'yov}
\email{solovyov@mbnresearch.com}
\affiliation{MBN Research Center, Altenh\"oferallee 3, 60438 Frankfurt am Main, Germany}

%\thanks{}
%\affiliation{}

\date{\today}

\begin{abstract}
This work reports on the development of a new approach to the multiscale computational modelling of the focused electron beam-induced deposition (FEBID), realised using the advanced software packages: MBN Explorer and MBN Studio. Our approach is based on stochastic dynamics (SD), which describes the probabilistic evolution of complex systems. The parameters for the new SD-based FEBID model were determined using molecular dynamics (MD) simulations. A new methodology was developed for this purpose and is described in detail. This methodology can be applied to many other case studies of the dynamics of complex systems. Our work focuses on the FEBID process involving W(CO)$_6$ precursor molecules deposited on a hydroxylated SiO$_2$-H substrate. Simulations and a detailed analysis of a growing W-rich nanostructure were performed. This new approach was shown to provide essential atomistic insights into the complex FEBID process, including the elemental composition and morphology of the deposit at each stage of growth. The derived results were then compared with experimental observations and validated. The multiscale methods developed in this study can be  further upgraded and applied to important technological developments, such as 3D nanoprinting.
\end{abstract}

\maketitle

%\interlinepenalty=10000
%\clubpenalty=10000
\widowpenalty=10000
%\brokenpenalty=10000
%%%%%%%%%%%%%%%%%%%%%%%%%%%%%%%%%%%%%%%%%%%%%%%%%%%%%%%%%%%%%
%%%%%%%%%%%%%%%%%%%%%%%%%%%%%%%%%%%%%%%%%%%%%%%%%%%%%%%%%%%%%
\section{Introduction}
\label{Intro}

\begin{figure}[t]
\centering
\includegraphics[width=13cm]{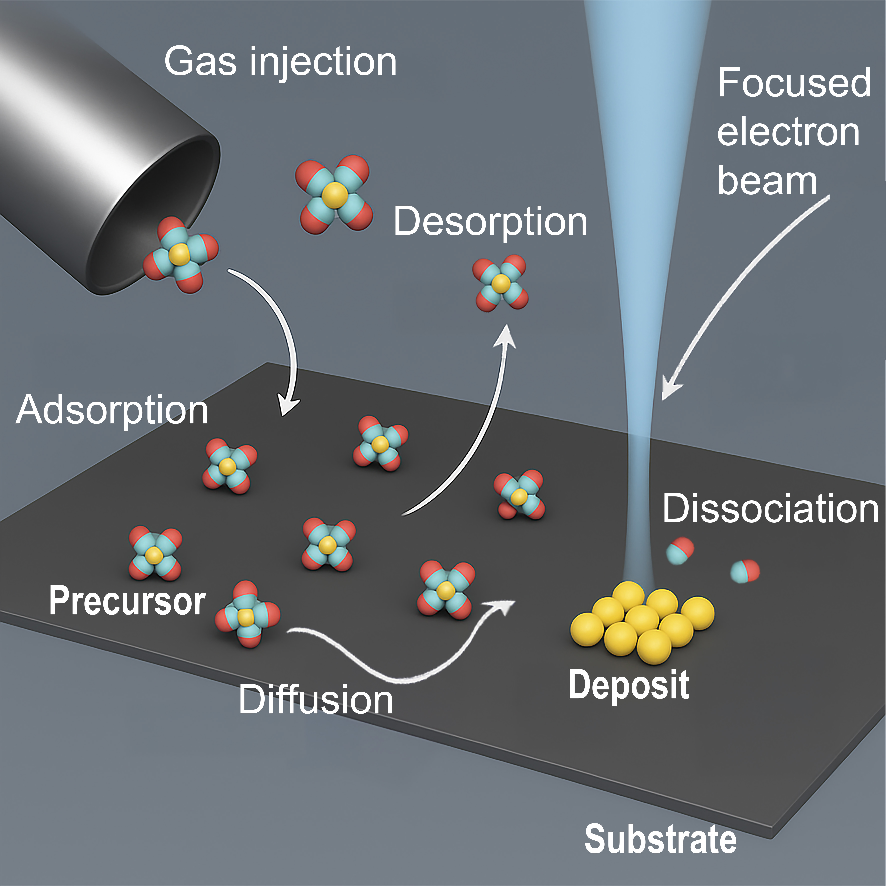}
\caption{A schematic of FEBID and the key elementary processes involved. A gas of precursor molecules is injected into the system and the molecules adsorb on the substrate surface. Through diffusion, the precursor molecules reach the irradiation area, where dissociation occurs. Precursor desorption from the substrate is also possible. The figure uses W(CO)$_6$ precursor molecules as an example, where the W-atom is shown in yellow, the C-atom in cyan and the O-atom in red. For illustration purposes, one of the CO-ligands is made transparent to better show the central position of the tungsten atom.}
\label{fig:FEBID}
\end{figure}

Focused electron beam-induced deposition (FEBID) is a modern technique for fabricating nanostructures with complex 3D geometries
\cite{Utke_book_2012, DeTeresa-book2020, Winkler2018,Keller2018}. In  FEBID, precursor molecules (mainly organometallic) are adsorbed onto a substrate surface and exposed to electron beam irradiation. While the volatile molecular fragments released during  the electron-induced decomposition of the precursors are pumped away, the remaining metal-containing fragments appear to be deposited on the surface, forming a structure with a characteristic size of its elements  determined by the incident electron beam (down to a few nanometers)\cite{Utke2008,Huth2012}. By changing the position of the electron beam, its current and the flux of deposited precursors a 3D structure of the metal deposit with the desired properties can be created.
A schematic of FEBID and the key elementary processes involved are shown in Fig.~\ref{fig:FEBID}.

FEBID typically operates through successive cycles of precursor molecule replenishment on a substrate and irradiation by a focused electron beam \cite{Utke2008,Huth2012}. This technique is governed by an interplay of many physical and chemical processes occurring at different characteristic temporal and spatial scales during the irradiation and replanishment phases of FEBID. For example, electron-induced precursor fragmentation occurs on the scale of femtoseconds. In contrast, the kinetic processes at the surface, such as adsorption, diffusion, and desorption, occur on much longer timescales of microseconds to milliseconds.

The critical technological challenge for the FEBID process is related to the current difficulty in controllable nanoscale fabrication of high-purity metal nanostructures of a desired geometry, size, and composition \cite{Keller2018,DeTeresa-book2020,Winkler2021,Huth_2021_JAP_review,Huth_2021_JApplPhys.130.170901}. Contamination of the nanostructures (typically with carbon and oxygen) is inherent in the FEBID process. The metal material produced by the FEBID process typically absorbs some of the precursor residues and other molecular species caused by the molecular fragmentation processes and associated chemistry \cite{Botman2009a}. Irradiation-driven chemistry is crucial in determining the grown structures' elemental composition and spatial resolution \cite{Sushko2016,Verkhovtsev2021}. There are other parasitic effects as well.  Thus, low-energy secondary electrons (SEs) emitted outside the focal point of the primary electron (PE) beam initiate electron-driven reactions and induce the formation of a halo, thus producing an undesired edge broadening of the structure \cite{Arnold2014}. To overcome these shortcomings, dedicated experimental, theoretical, and computational efforts have recently been undertaken to elucidate the underlying mechanisms of FEBID that determine the elemental composition and morphology of the grown deposits.

Apart from its fundamental value, the computational modeling approach to describe FEBID is seen as an attractive way to understand and improve the experimental procedures aimed at achieving a controlled growth of nanostructures with desired characteristics, see \cite{Sushko2016,Verkhovtsev2021} and references therein. 
Several approaches have been developed to model the FEBID process. Continuum models of FEBID are based on the reaction-diffusion differential equations, which describe the distribution of adsorbed precursor molecules on a substrate in terms of their concentration. These equations operate with the predefined rate constants \cite{Toth2015,Winkler2018}. Such an approach has been successfully used to model the growth rate of the nanostructures and their geometry. However, it often evokes empirical parameters and it  cannot provide detailed information about the  deposit's structure and its elemental composition. 

Atomistic structural and morphological features of the FEBID deposits can be described in detail using irradiation-driven molecular dynamics (IDMD) \cite{Sushko2016}. The IDMD methodology incorporates various quantum processes occurring in an irradiated system (e.g., covalent bond breakage induced by ionisation or dissociative electron attachment) into the classical MD framework, treating these processes stochastically  as random, fast, and local transformations of system \cite{Sushko2016}. This technique operates with the specially designed reactive CHARMM force field \cite{Sushko2016a}, for further details see \cite{Verkhovtsev2021} and references therein. The applicability of the IDMD methodology to the  simulation of the FEBID process has been explicitly demonstrated  for W(CO)$ _6 $ \cite{Sushko2016a, DeVera2020}, Pt(PF$ _3 $)$ _4  $\cite{Prosvetov2021}, Fe(CO)$ _5  $\cite{Prosvetov2021a}, Me$_2$Au(tfac) \cite{Prosvetov_Me2Autfac_2023} precursors deposited on the  SiO$_2$-H surface.  The IDMD approach provides a detailed atomistic description of all the processes involved in FEBID.  The advantage of this method lies in its ability to describe the elemental composition, morphology, as well as the growth rates of the fabricated structures. However, the spatial and temporal scales achievable in the atomistic IDMD simulations are typically limited to tens of nanometers and tens of nanoseconds due to the computational cost of the simulations.

This work presents an essential advancement in the multiscale description of the FEBID process that addresses the limitations of the atomistic IDMD approach.
It is  based on stochastic dynamics (SD), which describes the evolution of complex systems in a probabilistic manner by introducing  (i) a set of constituent elements/species populating the system, (ii) transformation processes between these species, and (iii) the rate constants for these processes. 

A versatile stochastic dynamics approach \cite{Solovyov2022} has been implemented within the universal MBN Explorer software package \cite{Solovyov2012}. The MBN Explorer software package is an advanced tool for multiscale simulations of the structure and dynamics of complex Meso-, Bio-, and Nano systems. It has been  successfully applied to  solve many different problems and case studies \cite{MBNbook_Springer_2017,Solovyov2024,DySoN_book_Springer_2022}. The software is suitable for classical non-relativistic and relativistic molecular dynamics (MD), Euler dynamics, reactive and irradiation-driven molecular dynamics (RMD and IDMD) simulations, as well as for stochastic dynamics, including kinetic Monte Carlo (KMC) simulations of various systems and processes relevant to physics, chemistry, biology, and materials science. The software enables a multiscale approach, linking different methodologies that describe dynamical phenomena on different temporal and spatial scales. In this work, MBN Explorer was used for the first time  to perform stochastic dynamics simulations of the FEBID process. The visualisation and analysis of the simulations was performed  using MBN Studio \cite{Sushko2019}. MBN Studio  is a special multitasking software toolkit with a graphical user interface developed for MBN Explorer that has been developed to facilitate setting up and starting MBN Explorer calculations, monitoring their progress and examining the calculation results. The MBN Explorer and MBN Studio, with their enormous potential for innovative solutions and multiscale computational modelling, formed the basis of the recent roadmap paper \cite{Solovyov2024}, which focused on advances in multiscale theory and simulation, and experimentation on condensed matter systems exposed to radiation. 

The stochastic dynamics approach implemented in MBN Explorer has been successfully applied to study the formation and evolution of nanofractlal stuctures created by deposition of metal clusters  on surfaces \cite{Solovyov2022,Solovyov_2014_PhysStatSolB.251.609, Panshenskov_2014_JCC.35.1317}, the stability of nanowires \cite{Moskovkin_2014_PSSB.251.1456} and the analysis of autocatalytic chemical reactions \cite{Solovyov2022}. Comparison of the simulation results with those of the related experiments carried out in these works demonstrated the method's effectiveness and  reliability in analysing complex dynamical processes beyond the temporal and spatial scales applicable to molecular dynamics methods.

This work presents the upper level multiscale approach for simulations of the FEBID process carried out using stochastic dynamics, whose parameters are determined by the molecular dynamics simulations. The parameters of the force fields used in the molecular dynamics are a result of the quantum theories (quantum chemistry and collision theory) derived in the previous work. This advanced multiscale approach takes into account all the elementary processes involved in FEBID mentioned above, such as adsorption, diffusion, desorption of the precursor molecules, their chemical reactions and radiation-induced fragmentation due to the interaction with the primary electron beam, secondary and backscattered electrons. The advantage of this method is that all these processes can be treated at their real times as they occur without rescaling and simulations can be carried out over a wide range of temporal and spatial scales, far beyond those typically accessible with molecular dynamics. Furthermore, this method allows all the essential atomistic insights into FEBID to be retained throughout these simulations. 

In such simulations the evolution of the system during FEBID is described through step-by-step transformations that occur on different particles within the system, such as intact and fragmented precursors, ligands, metal atoms, and the substrate. The physical and chemical transformations of the particles also include the precursor injection into the system, precursor diffusion, attachment, detachment, dissociation, and uptake. These are all discussed in detail in the following sections. For the stochastic description, these processes are defined by  characteristic probabilities per unit time interval evaluated according to the corresponding rate constants. The sections that follow discuss the derivation of the rate constants from atomistic molecular dynamics (MD) simulations of the underlying processes, as well as the track structure simulations for the incident electron beam.

The simulations performed in our work consider the FEBID of the W(CO)$_6$ precursor molecules on the SiO$_2$-H substrate with a 30 keV electron beam. This particular case study was chosen, because it has been studied experimentally in great detail \cite{Fowlkes2010,Geier_2014_JPCC.118.14009}  and can therefore be used as good reference case study, which has also been thoroughly analysed by  simulations using the atomistic IDMD approach introduced above \cite{Sushko2016,DeVera2020}.

In our work we demonstrate that SD simulations allow to reach temporal and spatial scales of the order of milliseconds and submicrometers, respectively, typical for the FEBID experiments \cite{Keller2018,DeTeresa-book2020,Winkler2021,Huth_2021_JAP_review,Huth_2021_JApplPhys.130.170901}. The verification of the simulation results is performed by a direct comparison with the experiment \cite{Fowlkes2010,Weirich2013,UTKE2022,Porrati2009,Huth2012}, where the experimental analysis of the average metal content, the size of deposits and their growth rates is carried out. It demonstrates that the simulation results are in good agreement  with the experimental data. Furthermore, they provide a highly detailed characterisation of the deposit's structure at a sub-nanometre level.

The established simulation protocol and its successful validation demonstrate that such SD simulations of the FEBID process and its analysis  can be carried out for a large number of other  case studies with many different precursors, substrates, irradiation and deposition regimes. Furthermore, this methodology can be further upgraded to perform simulations at even larger temporal and spatial scales relevant to FEBID-based 3D nanoprinting.

%%%%%%%%%%%%%%%%%%%%%%%%%%%%%%%%%%%%%%%%%%%%%%%%%%%%%%%%%%%%%%%%%%%
%%%%%%%%%%%%%%%%%%%%%%%%%%%%%%%%%%%%%%%%%%%%%%%%%%%%%%%%%%%%%%%%%%%
\section{Computational methodology}
\label{Methods}

\subsection{Stochastic dynamics: general aspects}

%Stochastic dynamics (SD) is useful for describing processes %that are probabilistic in nature.  Characterising the %properties of such processes may not require the knowledge of %the atomistic details of the species involved. It applies to %the modelling and analysis of systems composed of many %particles, where the constituent elements experience random %motion and may, for example, undergo various random %transformations and reactions \cite{Solovyov2022}. 
%The characterisation  of the properties of such processes may %not require the knowledge of the atomistic details of the %species involved. 
Stochastic dynamics (SD) is a method used to describe  systems that are  driven by probabilistic processes. The systems and processes involved can differ greatly  in nature.  The probabilistic  nature of motion  does not usually encompass all the degrees of freedom of the system. Instead, SD operates with a subset of constituent particles. These particles are then used to describe the system via the probabilistic processes involving them. For instance, SD dynamics of complex molecular systems may not require knowledge of the specific atomistic details of the constituent particles involved. Rather, knowledge of the probabilities of the processes involving these particles is required. These processes may involve the random motion of the particles or various random transformations and reactions involving them \cite{Solovyov2022}. 

%Therefore, SD can be a highly efficient method for modelling %such processes on the temporal and spatial scales beyond the %limits of molecular dynamics. Examples of such processes %include different modes of diffusion, dissociation and %attachment (depending on the system considered this can also %be decay, fission, and fusion), particle uptake and injection %(creation and annihilation), different reactive %transformations (e.g. chemical or collision induced), and %changes in particle types. The particles within these systems %can vary in nature, size, and properties, and interact with %each other in ways that affect their stochastic behaviour.
%%Understanding the associated dynamics often involves %considering hidden degrees of freedom and interparticle %interactions.

These facts explain  why SD is a highly efficient method of modelling the dynamics of complex molecular systems over the temporal and spatial scales that are beyond the limits of molecular dynamics. Examples of the processes involved in the SD dynamics  include diffusion, dissociation and attachment. Depending on the system in question, these processes may also include decay, fission, and fusion), particle uptake and injection (creation and annihilation),  different types of reactive transformation (e.g. chemical or collision-induced) and changes in particle types. The particles within such systems can vary in nature, size, and properties, and interact with each other in ways that affect their stochastic behaviour.

In the present study, we use SD to model the Focused Electron Beam Induced Deposition (FEBID) process of the W(CO)$_6$ precursor molecules atop the SiO$_2$-H substrate, demonstrating the method's ability to effectively capture the multiscale nature of the dynamic behaviour within this complex system. Below, we provide a brief description of the key ideas of the SD method while referring to the original methodological paper for a more detailed discussion \cite{Solovyov2022}. 
Furthermore, we use MD to derive the parameters necessary for SD modelling of the FEBID process.

\begin{figure}[t]
\centering
\includegraphics[width=15cm]{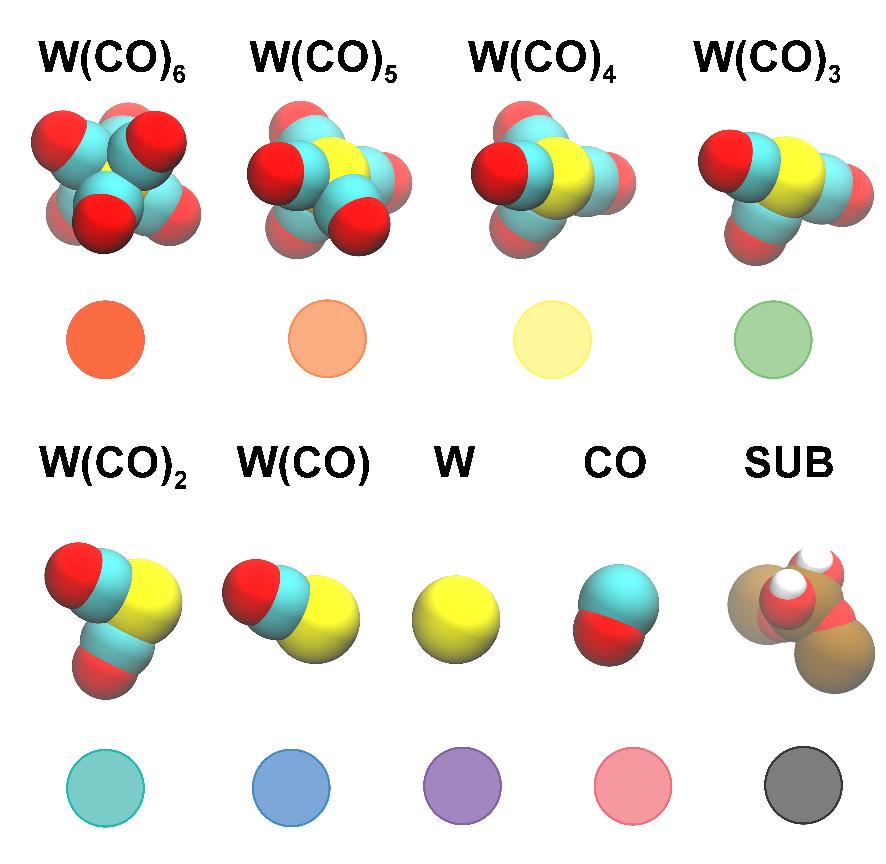}
\caption{Schematic representation of the structure of the precursor molecule W(CO)$_6$, its derivatives W(CO)$_{6-i}$, ($i=1,...,5$), metal W and the CO ligand. The precursor renderings show the atoms as spheres of the corresponding van der Waals radii. The circles below the molecules indicate the colouring scheme used to label the corresponding particles in the SD description of the FEBID process. The particle denoted as SUB denotes a fragment of the SiO$_2$-H substrate that size-wise roughly matches the size of one precursor molecule.}
\label{fig:introParticles}
\end{figure}

\begin{figure}[t]
\centering
\includegraphics[width=13cm]{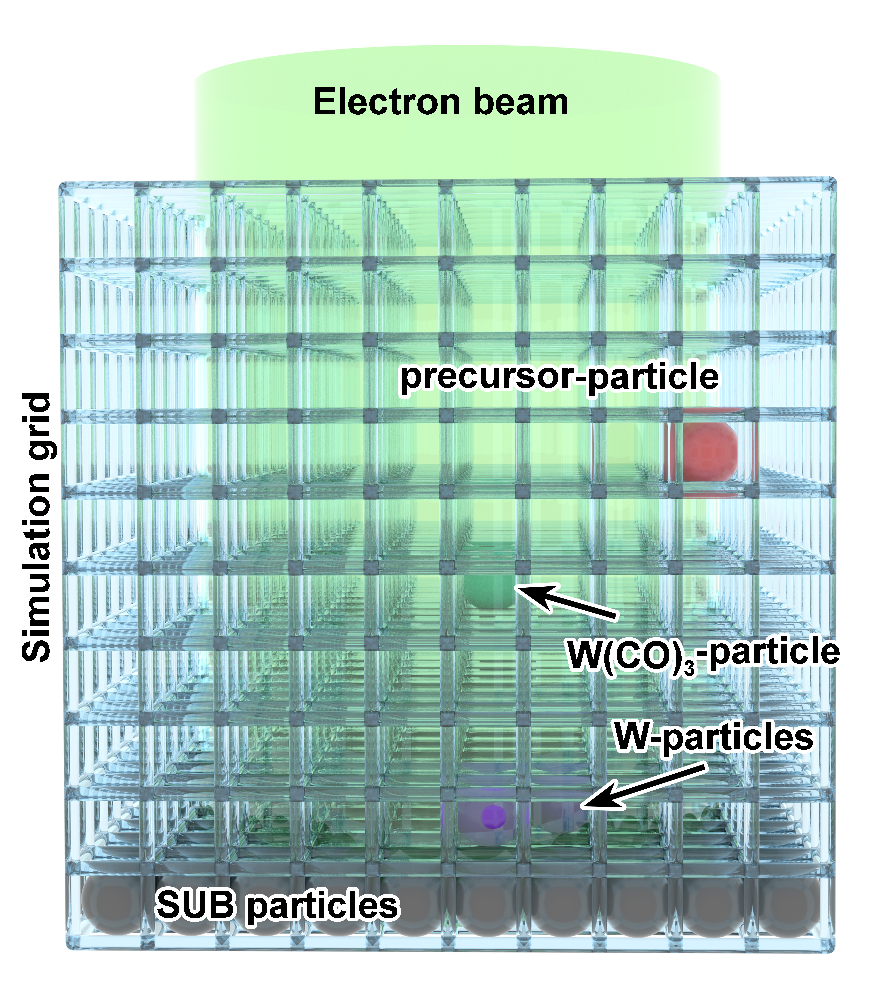}
\caption{Illustration of the three-dimensional simulation grid with particles of different types placed on it in different cells. The fixed layer of the particles at the bottom of the simulation box represents the substrate (SUB-particles). 
Precursor particles (red)  entering the area exposed to the electronic beam (green cylinder) undergo fragmentation. 
As an example, the products of such process (namely, a pure tungsten (W)-particle and a W(CO)$_{3}$ molecule) are shown.}
\label{fig:3Dgrid}
\end{figure}

In the SD description, the constituent particles are the fundamental elements of the system and are each characterised by a distinct type. The variety of processes within the system determines the number of particle types required for a simulation. The different particle types that are used to model the FEBID process of the W(CO)$_6$ precursor molecules atop the SiO$_2$-H surface are shown in Fig.~\ref{fig:introParticles}. For the present study it is essential to include the intact precursor molecules W(CO)$_6$, partially fragmented precursors with varying numbers of attached ligands W(CO)$_{6-i}$ (where $i=1,...,5$), pure metal atoms W, and detached ligand fragments CO. Additionally, the SiO$_2$-H substrate was modelled as a monolayer of immobile particles at the lower boundary of the simulation box, as illustrated in Fig.~\ref{fig:3Dgrid}. 
%The different particle types and their corresponding labels are introduced in Fig.~\ref{fig:introParticles}.
%\textcolor{red}{AVS: It is difficult to distinguish details in the black and white version.}

In SD, the studied FEBID process is modelled stepwise over time. At each simulation step, the system undergoes transformation according to the predefined probability of each possible process. Each new system configuration then serves as the starting point for the next simulation step, while the probabilistic transformations are governed by kinetic processes that involve all particles and occur at defined rates. These kinetic rates describe the probabilities of the process in the system happening per unit time. All the relevant ones are discussed in the following sections. The spatial evolution of particles during the FEBID process was modelled on a cubic grid where each particle's position is defined by three integer indices ($a$, $b$, $c$) corresponding to Cartesian coordinates ($x$, $y$, $z$), as illustrated in Fig.~\ref{fig:3Dgrid}. All particles are permitted to occupy individual grid cells and are transformed according to the predefined rules and probabilities, which may involve changes in particle properties (types) or movement to neighbouring vacant grid cells. The size of one grid cell is related to the physical size of the W(CO)$_6$ precursor molecule and was chosen as $d=0.63$ nm.
%The simulation box initially contained $ 51 \times 51 \times 27 $ unit cells, where the substrate particles occupied the lower layer, as illustrated in Fig.~\ref{fig:3Dgrid}. As the simulations progressed, the size of the simulation box was increased in the $z$-direction to allow the injection of new precursor particles into the system.

An important characteristic of SD is the simulation step $\Delta t$, which determines the time scale of the simulations. The time step is determined by the fastest process in the system and is related to the corresponding rate constant $\Gamma_0$ as \cite{Solovyov2022}:

\begin{equation}
\Gamma_0 \Delta t = 1.
\label{eq:dt}
\end{equation}

\noindent
Since all other processes in the system are expected to occur with lower rates, their respective probabilities $P_{\kappa}$ should be evaluated with the given time interval
$\Delta t$
%within an interval $[0...1]$ and can be evaluated 
as:

\begin{equation}
\label{key}
P_{\kappa} = \Delta t \Gamma_{\kappa},
\end{equation}

\noindent
where $\Gamma_{\kappa}$ is the rate constant of the corresponding process $\kappa$. SD description in MBN Explorer  permits various kinetic processes \cite{Solovyov2022}. In the present study, the following are relevant to model the FEBID process: (i) particle free displacement, (ii) particle peripheral diffusion, (iii) particle detachment, (iv) particle uptake, (v) particle injection, and (vi) particle dissociation. The naming of the kinetic processes here has been consistent with the original convention introduced earlier \cite{Solovyov2022}. To carry out the SD simulations of FEBID, the rate constants $\Gamma_{\kappa}$ in Eq.~(\ref{key}) should be determined for all the relevant processes in the system following the theoretical methodology described in \cite{Solovyov2022}. Below, we discuss in detail all the processes involved in FEBID and derive their rates from the atomistic analysis of the relevant properties of the precursor molecule and its fragments.
%Below, we explain how these processes are relevant to the FEBID process and how their %corresponding probabilities could be computed from the atomistic properties of the %precursor molecules and their fragments.

\subsection{Particle motion}

The FEBID process is modelled in multiple cycles consisting of two phases: (i) the irradiation phase and (ii) the replenishment phase. In simulations of each phase, all particles in the system are allowed to move stochastically. The motion of particles at each step is modelled through their translocation from one grid cell to the neighbouring ones. 
Three different translocation modes are considered: the displacement of free particles; the detachment of particles from each other and the diffusion of particles in different environments. 
The displacement of free particles, i.e. particles that have no contact with with other particles, is described by a  random jump to one of the neighbouring grid cells. The detachment process involves  a particle that was initially in contact with other particles moving to a neighbouring grid cell, whereby all the bonds between the detached particle and  neighboring particles in the system appearing to be broken.
Particle  diffusion describes the motion of a particle, where contact with a group of other particles in neighbouring grid cells remains maintained.

The diffusion of a particle at a grid cell defined by the three integer indices ($a$, $b$, $c$) depends on the particle type $\alpha$. The probability of this process is governed by the interplay of simultaneous breaking and maintaining of bonds between the moving particle and neighbouring particles. It can be factorised as \cite{Solovyov2022}:
%The probability of this process is governed by simultaneous breakage and maintenance of %bonds between particles and could be factorized as \cite{Solovyov2022}:

\begin{equation}
\label{Eq. P_per_jump_general}
P_{dif}^{\alpha}(a,b,c) =\sum_{\gamma=1}^{N_{open}}p_{jump}^{\gamma} \prod_{\beta=1}^{N_{types}} p_{dis}(\alpha, \beta)^{m(\alpha, \beta)} \prod_{\beta=1}^{N_{types}} p_{dif}(\alpha, \beta)^{n(\alpha, \beta)}.
\end{equation}
	
\noindent
Here $\alpha$ and $\beta$ denote two selected particle types in the system, while $N_{types}$ is the total number of particle types in the system. $p_{dis}(\alpha, \beta)$ is the probability of breaking a bond between particles of types $\alpha$ and $\beta$ (pairwise particle detachment probability), and $m$ is the total number of such bonds. $p_{dif}(\alpha,\beta)$ is the probability of the bond maintenance between a pair of particles of types $\alpha$ and $\beta$, while $n$ is the number of maintained bonds. The numbers $n$ and $m$ are specific for each position $(a,b,c)$ and each simulation step.  $p_{jump}^{\gamma}$ is the probability for a particle to jump to a neighbouring grid cell. These probabilities are not particle specific and are determined by geometry of the grid and the random walk principles \cite{Solovyov2022}. $N_{open}$ is the number of the open positions on the grid available for a jump from the point $(a,b,c)$ at a given simulation step. Equation~(\ref{Eq. P_per_jump_general}) is general and applicable to an arbitrary particle at any point on the simulation grid and at any simulation step. 

In the limiting case when all the bonds of a particle with neighbouring particles are broken, Eq.~(\ref{Eq. P_per_jump_general}) defines the probability of  a particle to detach from a group of other particles  \cite{Solovyov2022}.
%In the limiting case, Eq.~(\ref{Eq. P_per_jump_general}) can also be used to determine %the probability for a particle to detach from a group of other particles  %\cite{Solovyov2022}.  
In the most general case this group might consist of particles of different types and be arranged in a shape with an arbitrary geometry. The corresponding dissociation probability reads as

\begin{equation}
\label{Eq. P_detachment_general}
P_{dis}^{\alpha}(a,b,c) = \sum_{\gamma=1}^{N_{open}}p_{jump}^{\gamma}\prod_{\beta=1}^{N_{types}} p_{dis}(\alpha, \beta)^{m(\alpha, \beta)}.
\end{equation}

Equation~(\ref{Eq. P_per_jump_general}) is general and can be used for a large number of stochasticly moving systems, provided that the probabilities $p_{dis}(\alpha, \beta)$ and $p_{dif}(\alpha, \beta)$ are determined for each pair of particle types involved in the system dynamics. All other factors in this equation ($n$, $m$, $N_{open}$)  are  determined during simulations at each simulation step. They depend only on the configuration of particles in the vicinity of each grid point ($a$, $b$, $c$).

The probabilities  $p_{dis}(\alpha, \beta)$ and $p_{dis}(\alpha, \beta)$  in Eq.~(\ref{Eq. P_per_jump_general}) can be determined from atomistic molecular dynamics (MD) simulations. 
%aimed to probe the pairwise interactions between the precursor particles, their %fragments and the substrate (see below for the details of the MD simulations). 
To do this,  the probability $P_{dif}^{\alpha}(a,b,c)$ should be calculated using MD for a number of different processes and  geometries. The number of such cases should be equal to the number of the unknown probabilities $p_{dis}(\alpha, \beta)$ and $p_{dis}(\alpha, \beta)$ to be determined. This gives the sufficient number of nonlinear algebraic equations whose solutions determine the concrete values of $p_{dis}(\alpha, \beta)$ and $p_{dis}(\alpha, \beta)$ for each pair of particles involved in the motion. 

Below we apply this general method to FEBID, in which the diffusion of precursor molecules over a substrate and their detachment from it play a key role,
and determine the probabilities $p_{dif}(\alpha, \beta)$ and $p_{dis}(\alpha, \beta)$ 
of these processes for the case study considered in this paper.

\subsection{Precursor diffusion over and desorption from the substrate}

Let us use  Eq.~(\ref{Eq. P_per_jump_general}) to describe the diffusion of a precursor molecule over the substrate and Eq.~(\ref{Eq. P_detachment_general}) to describe its detachment from the substrate.
%Equations above are general and apply to arbitrary particle types and system geometries. 
%Let us now use these equations to consider a more specific case related to the problem %of FEBID and first discuss properties of precursors on the substrate. Eq.~(\ref{Eq. P_per_jump_general}) and 
In this case,  the types $\alpha$ and $\beta$ in Eq.~(\ref{Eq. P_per_jump_general}) and Eq.~(\ref{Eq. P_detachment_general}) should be more specific. For the sake of clarity, let us set $\beta=S$ to denote a particle of the substrate, while $\alpha_{0...6}$ will be used to denote the pure tungsten atom $\alpha_0$, an intact precursor $\alpha_6$ and precursor fragments $\alpha_{1...5}$.

According to  Eq.~(\ref{Eq. P_detachment_general}),
the probability of desorption (dissociation, detachment) of a particle from a substrate on the cubic grid, see Fig.~\ref{fig:3Dgrid}, can be calculated as:

\begin{equation}
\label{Eq. P_detachment_sub}
P_{dis}^{\alpha_i}(a,b)= \sum_{\gamma=1}^{N_{open}}p_{jump}^{\gamma}p_{dis}(\alpha_i, S)^{9}=N_{dis}p_{dis}(\alpha_i, S)^{9}.
\end{equation}

\noindent
Here $a$ and $b$ denote the coordinates of a grid cell on the substrate.  The ninth power of $p_{dis}(\alpha_i, S)$ in this equation arises because of a particle located just above the substrate on the cubic grid has nine neighbours from the substrate.  All the bonds with these neighbours become broken during the desorption process. In this case, $N_{dis} \approx 0.320$.  This factor determines the sum of the transition probabilities to the final states during the desorption (dissociation, detachment) process. It is determined by the geometry of the grid and the principles of random walk theory \cite{Solovyov2022}.

The probability of surface diffusion of a particle on the substrate can be derived from the general Eq.~(\ref{Eq. P_per_jump_general}). Considering this process on the cubic grid, the main mode of particle motion from one cell to a neighbouring cell on the substrate involves  breaking three bonds and the maintaining six bonds between the moving particle and  substrate particles. The resulting probability of the process can then be written as:

\begin{equation}
\label{Eq. P_diffusion_sub}
P_{dif}^{\alpha_i}(a,b) = \sum_{\gamma=1}^{N_{open}}p_{jump}^{\gamma}p_{dis}(\alpha_i, S)^3 p_{dif}(\alpha_i, S)^6=N_{dif}p_{dis}(\alpha_i, S)^3 p_{dif}(\alpha_i, S)^6.
\end{equation}

\noindent
Here, $N_{dif}\approx 0.357$ is the factor that determines the sum of the transition probabilities to the final states during the surface diffusion process. It is determined by the geometry of the grid and the principles of random walk theory \cite{Solovyov2022}.

The total probabilities $P_{dis}^{\alpha_i}(a,b)$ and  $P_{dif}^{\alpha_i}(a,b)$  can be derived from atomistic MD simulations, as shown in  the following subsections.
The pairwise probabilities $p_{dis}(\alpha_i, S)$ and $p_{dif}(\alpha_i, S)$  can then be obtained from equations Eqs.~(\ref{Eq. P_detachment_sub}) and (\ref{Eq. P_diffusion_sub}) by solving  them  for each given pair of $\alpha_i$ and $S$. These solutions read as

\begin{equation}
\label{p_dis}
p_{dis}(\alpha_i, S)= \left(\frac{P_{dis}^{\alpha_i}}{N_{dis}} \right)^{1/9},
\end{equation}

\begin{equation}
\label{p_dif}
p_{dif}(\alpha_i, S)= \left(\frac{P_{dif}^{\alpha_i}}{N_{dif}} \right)^{1/6}
\left(\frac{P_{dis}^{\alpha_i}}{N_{dis}} \right)^{-1/18}.
\end{equation}

Here we assume that the probabilities $P_{dis}^{\alpha_6} (a,b)$,  $P_{dif}^{\alpha_6} (a,b)$ do not depend on the coordinates of the precursor on the substrate. This assumption applies to the ideal,  defect-free substrate  considered in our work.
%otherwise these relations should be established for each point on the substrate.  

Now let us select the fastest process in the system. The desorption of precursor molecules from the substrate is usually a much slower process than their diffusion across the substrate, see e.g. \cite{Sushko2016}. Therefore, it is natural to expect that the pairwise probabilities $p_{dis}(\alpha_i, S)$ should be smaller than the corresponding probabilities $p_{dif}(\alpha_i, S)$. In the considered case study,  the intact precursor (the particle type $\alpha_6$)  should have the highest diffusion probability among the other movable particles in the system due to its weaker van der Waals interaction with the substrate. Therefore, putting $p_{dif}(\alpha_6, S)=1$, one can derive from Eqs. (\ref{p_dis}) and (\ref{p_dif}) the following expression for the time step for the SD simulations: 

\begin{equation}
\label{eq:dt2}
\Delta t=\frac{N_{dif}}{\Gamma_{dif}^{\alpha_6}}\left(\frac{N_{dif}}{\Gamma_{dif}^{\alpha_6}}\frac{\Gamma_{dis}^{\alpha_6}}{N_{dis}}\right)^{1/2}.
\end{equation}

Here $\Gamma_{dis}$ and $\Gamma_{dif}$ are defined through the following equations:
\begin{equation}
\label{eq:Gamma_dis}
P_{dis}^{\alpha_6}= \Delta t \Gamma_{dis}^{\alpha_6},
\end{equation}

\begin{equation}
\label{eq:Gamma_dif}
P_{dif}^{\alpha_6}= \Delta t \Gamma_{dif}^{\alpha_6} .
\end{equation}

\noindent 
Here we assume that also the rates of the processes $\Gamma_{dis}^{\alpha_6}$,  $\Gamma_{dif}^{\alpha_6}$  do not depend on the coordinates of the precursor on the substrate.
The probability of all other processes in the system can then be derived from Eq. (\ref{key}) with $\Delta t$ defined in Eq. (\ref{eq:dt2}).

Knowing $\Gamma_{dis}^{\alpha_6}$ and $\Gamma_{dif}^{\alpha_6}$, one can determine the time step $\Delta t$ for the SD simulations and the required probabilities $p_{dis}(\alpha_i, S)$, $p_{dif}(\alpha_i, S)$ used as the input parameters in MBN Explorer.

\subsection{Atomistic MD simulations of input parameters for SD}

The rates $\Gamma_{dis}^{\alpha_i}$ and $\Gamma_{dif}^{\alpha_i}$  and the probabilities  $P_{dis}^{\alpha_i}$ and $P_{dis}^{\alpha_i}$ can be derived from atomistic MD simulations using MBN Explorer \cite{Solovyov2012,Sushko2019,MBNbook_Springer_2017,Solo2017} for randomly oriented W(CO)$_i$ precursor molecules detaching from the SiO$_2$-H substrate surface, as well as for the pure W atom.
MD simulations can also be used to derive the diffusion and detachment rates for the $W(CO)_i$ species when they are in contact with neighbouring precursors.

\subsubsection{Rates of precursor detachment from the substrate}

The rates of detachment (desoption and dissociation) between different species (particles) can be derived from the analysis of the dependence of the interaction potential between them on the interparticle distance. This subsection is devoted to the analysis of such a dependence for (i) a randomly oriented W(CO)$_6$ precursor molecule detaching from the SiO$_2$-H substrate surface; and (ii) a W atom detaching from the SiO$_2$-H substrate surface. It should be noted that this method is general and can be applied to the analysis of many other molecular systems undergoing such processes.

The dependence of the interaction potential energy for two molecular species on the distance between them can be parametrised using the Morse potential as:

\begin{equation}
U(r_{\alpha\beta}) = D_{\alpha\beta} \left[ \exp\left[-2\lambda_{\alpha\beta}(r_{\alpha\beta} - r_0)\right] - 2\exp\left[-\lambda_{\alpha\beta}(r_{\alpha\beta} - r_0)\right] \right].
\label{Eq. Morse}
\end{equation}

\noindent
Here $D_{\alpha\beta}$ is the dissociation energy between particles of type $\alpha$ and $\beta$, $r_0$ is the equilibrium intermolecular distance, and $\lambda_{\alpha\beta} = \sqrt{k_{\alpha\beta} / 2 D_{\alpha\beta}}$ (with $k_{\alpha\beta}$ being the bond force constant) that determines the steepness of the potential.

The detachment rate for two species bound by such a potential can be evaluated using the Arrhenius approximation as:

\begin{equation}
\gamma(\alpha,\beta) = \nu_{\alpha\beta} \exp{\left(-\frac{D_{\alpha\beta}}{k_{\rm B} T}\right)}.
\label{Eq. Arrhenius}
\end{equation}

\noindent
Here $k_{\rm B}$ is the Boltzmann constant, and $T$ is the temperature, which in the calculations was assumed equal to 300~K. In the harmonic approximation, the vibration frequency $\nu_{\alpha\beta}$ of the particles at the minimum of the potential well in Eq.~(\ref{Eq. Morse}) read as

\begin{equation}
\nu_{\alpha\beta}=\frac{\lambda_{\alpha\beta}}{2\pi}\sqrt{\frac{2 D_{\alpha\beta}}{\mu_{\alpha\beta}}} = \frac{1}{2\pi}\sqrt{\frac{k_{\alpha\beta}}{\mu_{\alpha\beta}}}.
\label{Eq. AttempFrequency}
\end{equation}

\noindent
Here $\mu_{\alpha\beta}$ is the reduced mass of the particle pair $(\alpha\beta)$.

This theory can be used to evaluate the rates of the detachment (desorption,  dissociation) processes of different molecular species by approximating their actual interaction potentials with the Morse potential. In this case, the rate constant
for such processes can be evaluated as:

%Although atomistic MD simulations could be used to directly yield 
% the $p_{dis}(\alpha,\beta)$ probabilities in Eqs.~(\ref{Eq. P_per_jump_general})-
% (\ref{Eq. P_detachment_general}) it is  more advantageous 
%to calculate $P_{dis}^{\alpha}$ by modelling the dissociation process of a 
%particle $\alpha$ from a group of other particles. Such an approach will yield 
%an averaged probability of the detachment process while the direct calculation of 
%the pairwise probabilities $p_{dis}(\alpha,\beta)$ could lead to artifacts due to 
%the atomistic nuances of interparticle interactions at the interface. 
% The characteristic rate constant for the $P_{dis}^{\alpha}$ process is then computed as:

\begin{equation}
\Gamma_{dis}^{\alpha} = \tilde{\nu}_{\alpha} \exp{\left(-\frac{\tilde{D}_{\alpha}}{k_{\rm B} T}\right)}.
\label{Eq. Arrhenius2}
\end{equation}

\noindent
where $\tilde{\nu}_{\alpha}$ is the characteristic vibrational frequency of a pair of  molecular species bound together with each other by the interaction potential and $\tilde{D}_{\alpha}$ is their dissociation energy.
The rate constants derived with Eq.~(\ref{Eq. Arrhenius2}) can then be used in Eq.~(\ref{Eq. P_per_jump_general}) and Eqs. (\ref{eq:dt2}), (\ref{eq:Gamma_dis}),  (\ref{eq:Gamma_dif}) discussed above.

To derive the detachment rates $\Gamma_{dis}^{\alpha_i}$ for the W(CO)$_6$ precursor and its fragments from the SiO$_2$-H substrate, atomistic MD simulations have been performed to study  the dependence of the interaction potential energy of W(CO)$_6$ and a $W$ atom with the substrate on their distance from it. The details of the MD simulation protocol are provided below.

\begin{figure}[t]
\centering
\includegraphics[width=13cm]{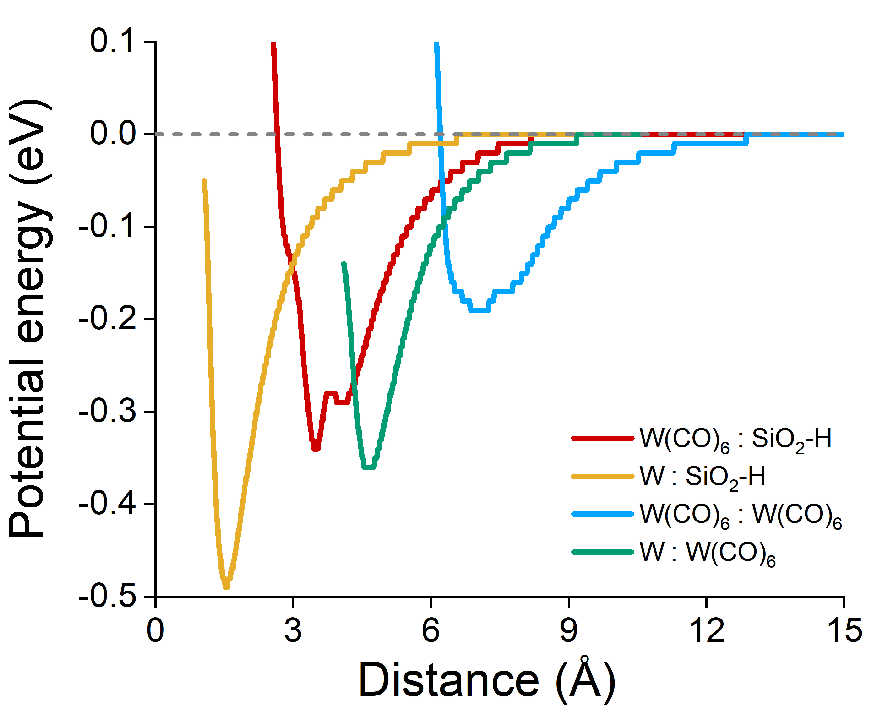}
\caption{ Dependence of the interaction potential energy of selected molecular species with the substrate on their distance from it. The distance is calculated from the centre of mass of the corresponding species. These dependencies are used for the derivation of parameters required for the SD simulations of FEBID carried out in this work.
The  dependences shown in red and in orange are  used to describe the detachment of the W(CO)$_6$ and W from the substrate, respectively, using Eq.~(\ref{Eq. Arrhenius2}). The dependencies shown in blue and in green are  used to describe the detachment of the W(CO)$_6$ and W from W(CO)$_6$ and W, respectively.}
\label{fig:AdsorptionEnergy}
\end{figure}

Figure~\ref{fig:AdsorptionEnergy} shows a number of such dependences, see e.g. red curve for a W(CO)$_6$ precursor molecule and orange curve for a $W$ atom both detaching from the SiO$_2$-H substrate surface. 
%of the interaction potential energies of selected molecular species (including substrate) on the distance between them obtained from atomistic MD simulations. 
%The selected W(CO)$_6$ precursor molecule (see red curve) and a W atom (see orange curve), both detaching from the SiO$_2$-H substrate surface, 
These dependencies are used for the derivation of parameters required for the SD simulations of FEBID carried out in this work.

Thus, the values of $\tilde{D}_{\alpha}$ and $\tilde{\nu}_{\alpha}$ in Eq.~(\ref{Eq. Arrhenius2}) are obtained for the W(CO)$_6$-substrate and W-substrate interactions by fitting the potential energy curves shown in Fig.~\ref{fig:AdsorptionEnergy} with the Morse potential defined in Eq.~(\ref{Eq. Morse}). The obtained detachment energies and the characteristic vibration frequencies have been averaged over the five simulations and are summarised in Table~\ref{tab:detachmentSUB}.

\begin{table}[htb]
\caption{Summary of the detachment parameters (detachment energies $\tilde{D}_{\alpha_i}$ and the characteristic vibrational frequencies $\tilde{\nu}_{\alpha_i}$) for precursor molecules and the tungsten atom detaching from the SiO$_2$-H  substrate derived from atomistic MD simulations. The presented parameters have been averaged over five independent MD simulations of the potential energy curves for the W(CO)$_6$ and $W$ atom interacting with  the substrate. The rate constants $\Gamma_{dis}^{\alpha_i}$ for W(CO)$_6$ and W atom have been evaluated  using Eq.~(\ref{Eq. Arrhenius2}).  The values of $\Gamma_{dis}^{\alpha_i}$ (i=1,..,5) for the fragmented precursors have been derived through the extrapolation between  the simulated values $\Gamma_{dis}^{\alpha_6}$ and $\Gamma_{dis}^{\alpha_0}$.}
\label{tab:detachmentSUB}
\centering
\begin{tabular}{p{1cm}p{3cm}p{2cm}p{2.5cm}p{2.5cm} }
    \toprule
	$i$ &	Particle ($\alpha_i$)         & $\tilde{D}_{\alpha_i}$ (eV)  & $\tilde{\nu}_{\alpha_i}$ (s$^{-1}$)    & $\Gamma_{dis}^{\alpha_i}$ (s$^{-1}$)\\
		\hline
	6	&   W(CO)$_6$                 &   0.33         & 5.24 $ \times  10^{11} $ &  1.67 $\times 10^6 $   \\
	5	&   	W(CO)$_5$                 &   -            & -                        &  8.26 $\times 10^5 $   \\
	4	&   	W(CO)$_4$                 &   -            & -                        &  3.81 $\times 10^5 $   \\
	3	&   	W(CO)$_3$                 &   -            & -                        &  1.68 $\times 10^5 $   \\
	2	&   	W(CO)$_2$                 &   -            & -                        &  7.18 $\times 10^4 $   \\
	1	&   	W(CO)                     &   -            & -                        &  3.00 $\times 10^4 $   \\
	0	&   	W                         &   0.48         & 1.65 $ \times  10^{12} $ &  1.24 $\times 10^4 $   \\
    \botrule
\end{tabular}
\end{table}

\begin{table}[t!]
\caption{Summary of the pairwise detachment probabilities $p_{dis}(\alpha_i, S)$ in Eq.~(\ref{p_dis}) of particles from the substrate. }
\centering
\begin{tabular}{p{1cm}p{3cm}p{2cm} }
    \toprule
    	$i$ & Particle ($\alpha_i$)  & \multicolumn{1}{c}{$p_{dis}(\alpha_i, S)$ }  \\
	\hline
       6 & W(CO)$_6$	  & \multicolumn{1}{c}{0.430}       \\  %0.430
       5 & W(CO)$_5$	  & \multicolumn{1}{c}{0.398}       \\  %0.398
       4 & W(CO)$_4$	  & \multicolumn{1}{c}{0.365}       \\  %0.365
       3 & W(CO)$_3$	  & \multicolumn{1}{c}{0.333}       \\  %0.333
       2 & W(CO)$_2$	  & \multicolumn{1}{c}{0.303}       \\  %0.303
       1 & W(CO)   	  & \multicolumn{1}{c}{0.275}       \\  %0.275
       0 & W	          & \multicolumn{1}{c}{0.250}       \\  %0.249
    \botrule
\end{tabular}
\label{tab:my_labelSUB}
\end{table}

The values of $\Gamma_{dis}^{\alpha_i}$ are summarised in Table~\ref{tab:detachmentSUB}. The values for the fragmented precursors have been derived through extrapolation between the simulated values $\Gamma_{dis}^{\alpha_6}$ and $\Gamma_{dis}^{\alpha_0}$.

In simulations, the detachment energies $\tilde{D}_{\alpha_i}$ and the vibrational frequencies $\tilde{\nu}_{\alpha_i}$ have been linearly extrapolated between the values obtained for an intact precursor and a single $W$ atom as a function of the number of detached ligands. The resulting values have then been substituted in Eq.~(\ref{Eq. Arrhenius2}) to derive all the relevant $\Gamma_{dis}^{\alpha_i}$ values. 
%The values of $\Gamma_{dis}^{\alpha_i}$  obtained here %correspond to a zero-order approximation for their evaluation %and should be correct by the order of magnitude. 
Further refinement of this approach may account for the non-harmonic behaviour of the interaction potential energy curves and go beyond the Arrhenius approximation, see Eq.~(\ref{Eq. Arrhenius2}).

%, that were used to calculate the partial probabilities $p_b(i, j)$ and $p_{\mu}(i, j)$ in Eqs.~(\ref{Eq. P_detachment_general})-(\ref{Eq. P_diffusion_sub})

The probabilities for the different pairwise dissociation processes used in the simulations are compiled in Table~\ref{tab:my_labelSUB}. They have been derived using the values of $\Gamma_{dis}^{\alpha_i}$ from Table~\ref{tab:detachmentSUB} and Eq.~(\ref{p_dis}), where $P_{dis}^{\alpha_i}$ is obtained from Eq.~(\ref{eq:Gamma_dis}). The calculation of $p_{dis}(\alpha_i, S)$ also requires $\Delta t$, defined in Eq.~(\ref{eq:dt2}), which is determined by $\Gamma_{dif}^{\alpha_6}$ and $\Gamma_{dis}^{\alpha_6}$. Using the values for $\Gamma_{dis}^{\alpha_6}$ from Table~\ref{tab:detachmentSUB} and $\Gamma_{dif}^{\alpha_6}$ determined below, one obtains $\Delta t=0.097$ ns.

\subsubsection{Rates of precursor diffusion across the substrate}

The surface diffusion of (i) a W(CO)$_6$ precursor molecule and a $W$-atom  over the SiO$_2$-H substrate can also be performed by means of atomistic MD simulations in order to derive the corresponding diffusion coefficients. With their use,  $P_{dif}^{\alpha_6}$ and $P_{dif}^{\alpha_0}$ probabilities in Eqs.(\ref{Eq. P_diffusion_sub}) and (\ref{p_dif}) can be set up. The details of the related analysis 
%MD simulation protocol 
are provided below.

To determine the partial probabilities $p_{dif}(\alpha_i, S)$ in Eq.~(\ref{p_dif}) it is necessary to know the diffusion coefficients of W(CO)$_6$ precursor and a  $W$-atom atop the SiO$_2$-H substrate. These values have been adapted from the literature, see  \cite{Solovyov2012,Sushko2019,MBNbook_Springer_2017, DySoN_book_Springer_2022,Solo2017,Solo2024}, and read as $\mathcal{D}^{W6}=29.10$ $\mu m^2/s$ and $\mathcal{D}^{W0}=0.21$ $\mu m^2/s$. These numbers have been used for the evaluation of the corresponding surface diffusion rate constants for a W(CO)$_6$ precursor and a  $W$-atom atop the SiO$_2$-H substrate, resulting in $\Gamma_{dif}^{\alpha_6}=2.93\times10^8$ 1/s and $\Gamma_{dif}^{\alpha_0}=2.15\times10^6$ 1/s. These rate constants determine the probability $P_{dif}^{\alpha_i}$ in Eqs.~(\ref{Eq. P_diffusion_sub}), ~(\ref{p_dif}) and ~(\ref{eq:Gamma_dif}).

\begin{table}[b!]
\caption{Summary of the pairwise probabilities $p_{dif}(\alpha_i, S)$ describing the translocation of different particles on the substrate used in the SD simulations of the FEBID process.}
\centering
\begin{tabular}{p{1cm}p{3cm}p{2cm}}
    \toprule
    $i$	& Particle ($\alpha_i$) &  $p_{dif}(\alpha_i, S)$   \\
	\hline
      6 &  W(CO)$_6$	  & 1.000       \\  %1.0
      5 &  W(CO)$_5$	  & 0.930       \\  %0.925
      4 &  W(CO)$_4$	  & 0.860       \\  %0.848
      3 &  W(CO)$_3$	  & 0.789       \\  %0.774
      2 &  W(CO)$_2$	  & 0.719       \\  %0.704
      1 &  W(CO)    	  & 0.649       \\  %0.639
      0 &  W            & 0.579       \\  %0.579
    \botrule
\end{tabular}
\label{tab:Pmu}
\end{table}

As explained above, the diffusion of the W(CO)$_6$ precursor over the surface occurs with a unit probability (as it is the fastest process in the system) during one simulation step, i.e. $p_{dif}(\alpha_6, S)=1$.

Using the values of $\Gamma_{dif}^{\alpha_0}$ and $\Gamma_{dis}^{\alpha_0}$ obtained from the MD simulations and presented  in Table~\ref{tab:detachmentSUB}), one derives from Eq.~(\ref{p_dif}) the value $p_{dif}(\alpha_0, S)=0.579$. The probabilities of the diffusion of particles describing the precursor, the tungsten, and precursor fragments over the substrate are summarised in Table~\ref{tab:Pmu}. The diffusion probabilities for precursor fragments  W(CO)$_i$ (i=1...5) over the substrate have been  derived through a linear interpolation of the $p_{dif}(\alpha_6, S)$ and $p_{dif}(\alpha_0, S)$ values.

%\subsection{Properties of precursors and their fragments}
\subsubsection{ Self diffusion and detachment rates of precursors and their fragments}

Now let us determine
the rates of self diffusion and detachment of precursors and their fragments. These rates could be evaluated similarly to those discussed above for the detachment and diffusion of particles on the surface. Thus, the rates of detachment (desorption and dissociation) between different species (particles) can be derived from obtained from MD simulations,  analysing the dependence of the interaction potential energy upon the interparticle distance. This subsection is devoted to the analysis of such dependences for (i) a randomly oriented W(CO)$_6$ precursor molecule interacting with a cluster of W(CO)$_6$ molecules; (ii) a W-atom interacting a cluster of W(CO)$_6$ molecules. Examples of the simulated  dependences are shown in Fig.~\ref{fig:AdsorptionEnergy}.

%The partial probabilities $p_{dis}(\alpha,\beta)$ and $p_{dif}(\alpha,\beta)$ in Eqs.~(\ref{Eq. P_per_jump_general})-(\ref{Eq. P_detachment_general}) for precursor particles and their fragments were established from the atomistic MD simulations, where the pairwise interactions between the particles were probed. MD simulations were performed using MBN Explorer \cite{Solovyov2012,Sushko2019,MBNbook_Springer_2017,Solo2017} for two scenarios: (i) a randomly oriented W(CO)$_6$ precursor molecule approaching a cluster of W(CO)$_6$ molecules; (ii) a W-atom approaching a cluster of W(CO)$_6$ molecules. In the following, $\alpha_{0...6}$ will be used instead of $\alpha$ and $\beta$ to denote the pure tungsten atom $\alpha_0$, an intact precursor $\alpha_6$ and precursor fragments $\alpha_{1...5}$, as also done previously. 

Following the methodology outlined above, the dependences of the interaction potential energies on the distance between the particles have been used to determine $\Gamma_{dis}^{\alpha_6}$ and $\Gamma_{dis}^{\alpha_0}$ for W(CO)$_6$ and W particles detaching from a cluster of W(CO)$_6$ particles. For this purpose, first the values of $\tilde{D}_{\alpha}$ and $\tilde{\nu}_{\alpha}$, see Eq.~(\ref{Eq. Arrhenius2}), for the (i) W(CO)$_6$ : W(CO)$_6$, and (ii) W : W(CO)$_6$ interactions have been obtained from five independent MD simulations. The corresponding averaged values of $\tilde{D}_{\alpha}$ and $\tilde{\nu}_{\alpha}$ are compiled in Table~\ref{tab:detachmentPREC}. The characteristic rate constants $\Gamma_{dis}^{\alpha_i}$ for the  detachment processes of the fragmented precursors W(CO)$_i$  have been derived through the linear extrapolation from the values 
$\Gamma_{dis}^{\alpha_6}$  and  $\Gamma_{dis}^{\alpha_0}$.

\begin{table}[!t]
\caption{Summary of the averaged parameters (the detachment energies $\tilde{D}_{\alpha_i}$ and the characteristic vibration frequencies $\tilde{\nu}_{\alpha_i}$) for the precursor molecule W(CO)$_6$ and its fragments, including the tungsten atom, derived from atomistic MD simulations for  the detachment of the molecular species from an island of W(CO)$_6$ precursors. Averaging has been performed over five independent MD simulations. The detachment rate constants $\Gamma_{dis}^{\alpha_i}$ have been evaluated using Eq.~(\ref{Eq. Arrhenius2}) for W(CO)$_6$ and W species, for which  the interaction energy dependencies have been simulated.  For the fragmented precursors, the detachment rate constant have been derived through extrapolation between the simulated values.}
\label{tab:detachmentPREC}
\centering
\begin{tabular}{p{1cm}p{3cm}p{2cm}p{2.5cm}p{2.5cm} }
    \toprule
		$i$  &  Particle (\textit{$\alpha_i$})   &  $\tilde{D}_{\alpha_i}$ (eV)  & $\tilde{\nu}_{\alpha_i}$ (s$^{-1}$)    & $\Gamma_{dis}^{\alpha_i}$ (s$^{-1}$)\\
		\hline
		6  &  W(CO)$_6$                  &   0.21          & 6.08 $ \times  10^{11} $ &  2.08 $\times 10^8 $   \\
		5  &  W(CO)$_5$                  &   -             & -                        &  7.80 $\times 10^7 $   \\
		4  &  W(CO)$_4$                  &   -             & -                        &  2.74 $\times 10^7 $   \\
		3  &  W(CO)$_3$                  &   -             & -                        &  9.22 $\times 10^6 $   \\
		2  &  W(CO)$_2$                  &   -             & -                        &  3.02 $\times 10^6 $   \\
		1  &  W(CO)                      &   -             & -                        &  9.66 $\times 10^5 $   \\
		0  &  W                          &   0.40          & 1.86 $ \times  10^{12} $ &  3.05 $\times 10^5 $   \\
    \botrule
\end{tabular}
\end{table}

The partial probabilities $p_{dis}(\alpha_i,\alpha_6)$ have been derived using Eq.~(\ref{p_dis}) with  $P_{dis}^{\alpha_i}$ calculated with the $\Gamma_{dis}^{\alpha_i}$ values collected in Table~\ref{tab:detachmentPREC}. The derived values for the  particle detachment probabilities can be refined in many different ways.  The vibration frequencies of particles can be  evaluated beyond the harmonic approximation, the fragmentation process can be treated beyond the Arrhenius approximation, the detachment rates for different combinations of fragmented precursors can be evaluated more accurately.  However, all these improvements would require extended MD calculations and conceptually would not add much to the main focus of the present study. Therefore, instead of dwelling into parameter development, in addition to the set of partial probabilities $p_{dis}(\alpha_i,\alpha_6)$ evaluated directly from on the $\Gamma_{dis}^{\alpha_i}$ values presented in Table~\ref{tab:detachmentPREC} (Set 1), we have also considered two additional sets of partial probabilities $p_{dis}(\alpha_i,\alpha_6)$ (Set 2 and Set 3).

%All partial probabilities of precursor particles, tungsten particles and their fragments detachment from the pure precursor particle W(CO)$_6$, employed in the simulations, are compiled Table~\ref{tab:my_labelPREC}. 
Table~\ref{tab:my_labelPREC} summarises the partial probabilities  derived directly from Eq.~(\ref{p_dis}) for the  Set 1, using the values of $\Gamma_{dis}^{\alpha_i}$  from Table~\ref{tab:detachmentPREC}. The parameters for Sets 2 and 3 were  generated from those for  Set 1, with the pairwise probabilities of a pure precursor detaching from its fragmented variants scaled by factors of 2.065 and 1.460, respectively.
This was done to explore the influence of these parameters on the observable characteristics of the FEBID process.

\begin{table}[t!]
\caption{A summary of the pairwise detachment probabilities $p_{dis}(\alpha_i, \alpha_6)$, computed using  Eq.~(\ref{p_dis}) for W(CO)$_6$ precursor, a tungsten atom and the precursor fragments. Three sets of probabilities (Set 1, Set 2, Set 3) have been used to in the FEBID simulations presented below.   }
\centering
\begin{tabular}{p{1cm}p{3cm}p{2cm}p{2cm}p{2cm} }
    \toprule
    	$i$ & Particle ($\alpha_i$)  & \multicolumn{3}{c}{$p_{dis}(\alpha_i, \alpha_6)$ }  \\
    	{}      &      & Set 1     & Set 2    & Set 3    \\
	\hline
       6 & W(CO)$_6$	  &  \multicolumn{3}{c}{$\leftarrow$0.735$\rightarrow$}       \\  %0.706 ???
       5 & W(CO)$_5$	  &  0.660     & 0.319    & 0.452    \\  %0.638     & 0.319    & 0.451   %run 47 | run 46 | run 48
       4 & W(CO)$_4$	  &  0.587     & 0.284    & 0.402    \\  %0.570     & 0.285    & 0.403
       3 & W(CO)$_3$	  &  0.520     & 0.252    & 0.356    \\  %0.506     & 0.253    & 0.358
       2 & W(CO)$_2$	  &  0.460     & 0.223    & 0.315    \\  %0.448     & 0.224    & 0.317
       1 & W(CO)   	  &  0.405     & 0.196    & 0.277    \\  %0.395     & 0.198    & 0.280
       0 & W	          &  0.356     & 0.173    & 0.244    \\  %0.347     & 0.174    & 0.246
    \botrule
\end{tabular}
\label{tab:my_labelPREC}
\end{table}

To approximate the pairwise diffusion probabilities $p_{dif}(\alpha_i, \alpha_j)$ for different precursor fragments relative to each other, we  used the established pairwise probabilities, $p_{dif}(\alpha_i, S)$, describing the translocation of different particles on the substrate (see Table~\ref{tab:Pmu}). The probabilities $p_{dif}(\alpha_i, \alpha_j)$ used  in this study are compiled in Table~\ref{tab:prob_pairwise}.
%(i)   $p_{dif}(\alpha_6, \alpha_{0\ldots6})=p_{dif}(\alpha_6, S)=1$;\\
%(ii)  $p_{dif}(\alpha_5, \alpha_{0\ldots5})=p_{dif}(\alpha_5, S)=0.930$;\\
%(iii) $p_{dif}(\alpha_4, \alpha_{0\ldots4})=p_{dif}(\alpha_4, S)=0.860$;\\
%(iv)  $p_{dif}(\alpha_3, \alpha_{0\ldots3})=p_{dif}(\alpha_3, S)=0.789$;\\
%(v)   $p_{dif}(\alpha_2, \alpha_{0\ldots2})=p_{dif}(\alpha_2, S)=0.719$;\\
%(vi)  $p_{dif}(\alpha_1, \alpha_{0\ldots1})=p_{dif}(\alpha_1, S)=0.649$;\\
%(vii) $p_{dif}(\alpha_0, \alpha_0)=p_{dif}(\alpha_0, S)=0.579$.\\
%\textcolor{red}{The values above are consistent with Table~\ref{tab:Pmu}, but are slightly different from the actual simulation file. Also $p_{dif}(\alpha_6, \alpha_1)=0.918$ and $p_{dif}(\alpha_6, \alpha_0)=0.808$ in the input file, and I have no idea why. In the text above, I set them to 1.0}

\begin{table}[h]
    \centering
    \begin{tabular}{p{1.5cm}|p{1.5cm}p{1.5cm}p{1.5cm}p{1.5cm}p{1.5cm}p{1.5cm}p{1.5cm}}
    \toprule
     \diagbox[width=1.5cm]{$\alpha_i$}{$\alpha_j$} & $\alpha_6$ & $\alpha_5$ & $\alpha_4$ & $\alpha_3$ & $\alpha_2$ & $\alpha_1$ & $\alpha_0$ \\ \hline
    $\alpha_6$ & 1.000 & 1.000 & 1.000 & 1.000 & 1.000 & 1.000 & 1.000 \\ 
    $\alpha_5$ & 1.000 & 0.930 & 0.930 & 0.930 & 0.930 & 0.930 & 0.930 \\ 
    $\alpha_4$ & 1.000 & 0.930 & 0.860 & 0.860 & 0.860 & 0.860 & 0.860 \\ 
    $\alpha_3$ & 1.000 & 0.930 & 0.860 & 0.789 & 0.789 & 0.789 & 0.789 \\ 
    $\alpha_2$ & 1.000 & 0.930 & 0.860 & 0.789 & 0.719 & 0.719 & 0.719 \\ 
    $\alpha_1$ & 1.000 & 0.930 & 0.860 & 0.789 & 0.719 & 0.649 & 0.649 \\ 
    $\alpha_0$ & 1.000 & 0.930 & 0.860 & 0.789 & 0.719 & 0.649 & 0.579 \\ 
    \hline
    $S$        & 1.000 & 0.930 & 0.860 & 0.789 & 0.719 & 0.649 & 0.579 \\
    \botrule
    \end{tabular}
    \caption{The pairwise probabilities, $p_{dif}(\alpha_i, \alpha_j)$, describe the relative diffusion of particles of different types. These probabilities are consistent with the values presented in Table~\ref{tab:Pmu}. 
    %Pairwise probabilities $p_{dif}(\alpha_i, \alpha_j)$ describing the relative diffusion of different particle types. These probabilities were taken to be consistent with the $p_{dif}(\alpha_i, S)$ probabilities describing the translocation of different particles on the substrate, compiled in Table~\ref{tab:Pmu}.
    }
    \label{tab:prob_pairwise}
\end{table}

The above probabilities were used to model the different  processes occurring between particles in contact.  In simulations, particles can also be moved to neighbouring grid cells if they are not in contact with any other particles.
Such situation arises  when new precursors  are  injected into the system,  or when deposited molecules on the surface fragment into the bulk. The motion of free particles in the bulk of the simulation box can be less stochastic; for example, when particles (i.e. molecules) in a flow  move towards the substrate. However, for FEBID modelling, the details of such dynamics are not so essential. The only  characteristic relevant to the modelling is the flux of molecules from the gas that reaches the substrate. Therefore, to  simplify the simulation protocol, we modelled the motion of molecules in the gas within the SD framework, assuming a unit probability of translocation of such  molecules  in the bulk at every simulation step.

The Arrhenius equation Eq.~(\ref{Eq. Arrhenius2}) provides a relatively simple method of estimating the rate of molecular fragmentation. A more accurate treatment of the chemical reactivity of molecules could further stabilise their binding and lead to lower detachment rates. In the simulations,  we assumed that the detachment probability of any two fragmented precursors (including pure tungsten) was zero. This assumption is justified by the strong metallic bond formed between metal atoms when two fragmented precursors come into contact.

Conversely, the detachment probability of the CO ligand particles from all other particles was set to 1, because the CO ligand species (see Fig.~\ref{fig:introParticles}) were assumed to be volatile and inert once formed. They were therefore not considered in any possible follow-up reactions.  Consistent with this approach, the pairwise diffusion probability of  CO particles with any other particles in the system was also set to 1.

\subsection{Addition of particles}

The injection of new precursor molecules  into the system was modelled in accordance with the experimental data.  In the SD simulations, the molecular flux of the precursor molecules impinging on the surface was used to calculate the probability of new particles being injected  into the  bulk. According to the kinetic theory of gases, the uniform  flux $F_{\rm p}$ of molecules landing on a surface in a chamber with pressure $P_{\rm p}$ is given by \cite{LANDAU1980111}:

\begin{equation}
F_{\rm p} = \frac{P_{\rm p}}{\sqrt{2\pi \, m_{\rm p} \, k_{\rm B} T_{\rm p}}},
\label{Eq. GasFlux}
\end{equation}

\noindent
where $T_{\rm p}$ is the gas temperature, $m_{\rm p}$ is the mass of the precursor molecule, and $k_{\rm B}$ is the Boltzmann constant. The molecular flux of the W(CO)$_6$ precursors, calculated using Eq.~(\ref{Eq. GasFlux}) at 300~K and a gas pressure of $P=0.333$ Pa \cite{Fowlkes2010}, is equal to $F_p = 2700$ molecules/(nm$^2$s). In the SD FEBID simulations, the precursor injection rate  was calculated so that the number of newly added precursors in the simulation box equalled the number of impinging precursor molecules within the replenishment time interval. The latter number can be calculated as follows

\begin{equation}
\label{eq:particlesToadd}
N_{add}=F_p S T_{rep}=F_p N_x N_yd^2 N_{rep}\Delta t,
\end{equation}

\noindent
where $S$ is the substrate surface area on which the particles  land, $T_{rep}$ is the replenishment time, $N_x$ and $N_y$ are the number of grid cells in the $x$- and $y$-directions that comprise the available  substrate  area for the new particles, $d$ is the size of one grid cell and $N_{rep}$ is the number of simulation steps during the replenishment phase. In the simulations performed, $N_x=N_y=51$ and $N_{rep}=1.28\times10^6$, while $d=0.63$ nm.  With these values, $N_{add}=346$ particles.

In the simulations, the newly added precursor molecules were  randomly placed within the empty grid cells of the simulation box. The particles were then permitted to move stochastically within the simulation box. However, this type of motion does not accurately represent that happens in a real experiment, where molecules are usually directed towards the substrate. A more accurate  modelling of such dynamics would not only complicate the description of the problem, but would also be unnecessary for our purposes.
Instead, the SD simulations were carried out so that the number of molecules reaching the substrate surface would corresponded roughly to the theoretical value in Eq.~(ref{eq:particlesToadd}). To prevent the precursor molecules from clustering in the bulk of the simulation box, absorbing boundary conditions were introduced to the upper boundary of the simulation box to remove all particles that reached it.

The probability of adding a new particle addition at each simulation step was set to $P_a=0.0157$, resulting in an average number $N_{add}$ of precursor molecules being absorbed onto the surface during the replenishment  interval, in line  with Eq.~(\ref{eq:particlesToadd}). Precursor molecules were continuously injected into the system at every simulation step with the probability $P_a$  at every simulation step during the irradiation and replenishment phases.

\subsection{Fragmentation of particles}

Electron-induced fragmentation of precursor molecules is another process that plays an important role in FEBID. There are several types of collisions that lead to precursor fragmentation. At electron collision energies below ~10 eV, the dissociative electron attachment and dissociative electronic excitations 
are the dominant processes, whereas at higher collision energies, dissociative ionisation becomes dominant, see e.g. \cite{DeVera2020}. 

In SD, electron-induced fragmentation of precursor molecules was described by particle dissociation, which occurs with some  probability in the irradiated part of the system. In this case, at each  step of the simulation, each irradiated particle can with some probability be replaced  by two particles, namely a precursor with one ligand less and a detached ligand. The fragmentation rate of the precursors was calculated from the spatial distribution of the flux density $J$ of primary (PE), secondary (SE), and backscattered electrons (BSE)  and the absolute fragmentation cross section of the precursor molecules $\sigma_{\rm frag}$ \cite{Sushko2016,DeVera2020}:

\begin{equation}
P(x,y) = \sigma_{\rm frag}(E_0) J_{\rm PE}(x,y,E_0) + \sum_i \sigma_{\rm frag}(E_i) [J_{\rm SE}(x,y,E_i) + J_{\rm BSE}(x,y,E_i) ].
\label{Eq. Frag_Probability_total}
\end{equation}

\noindent
Here $\sigma_{\rm frag}(E)$ is the energy-dependent precursor fragmentation cross section, $E_i < E_0$ is the electron energy discretized in steps of 1~eV, $E_0$ is the primary electron energy and $J_{\rm PE/SE/BSE}(x,y,E_i)$ are the flux densities of PE, SE, and BSE with energies $E_i$ taken at the point ($x$, $y$). 
The primary electrons are the electrons from the electron beam.  The secondary electrons are the electrons emitted due to the ionisation of the substrate (dominant fraction), but also from the ionisation of deposited precursors. The backscattered electrons are the electrons from the primary beam that have been backscattered by elastic collisions with the nuclei of atoms, mostly in the substrate.
The electron flux density can be evaluated from the experimental conditions using the values of the electron current $I$ and the beam spot radius $R_{\rm b}$ as $J_{\rm PE}=I /( e \pi R_{\rm b}^2)$.

The fragmentation rate distribution of W(CO)$_6$ irradiated by the 30 keV electron beam is taken from the earlier study \cite{DeVera2020}. 
There, the rate constants of the electron-induced dissociation of precursor molecules were determined  using the fragmentation cross section and the spatial distribution of the electron flux density obtained with the track-structure Monte Carlo code SEED, see \cite{DeVera2020} for further details and references therein. Here, for simulation efficiency, this distribution was fitted with a piecewise function to describe the radial dependence of the fragmentation rate distribution both within and outside of the primary beam spot area as

\begin{equation}
\label{Eq:BeamProfile}
\left\{\begin{array}{@{}l@{}}
P = A + B *\left[ 1 - 1/(1+e^{-(|r-r_c|-R_b)/w_0})  \right]/\left(1+e^{-(|r-r_c|+R_b)/w_0}\right), \ \ \ |r-r_c| \le R_b , \\
P = B *\left[ 1 - 1/(1+e^{-(|r-r_c|-R_b)/w_0})  \right]/\left(1+e^{-(|r-r_c|+R_b)/w_0}\right), \ \ \ |r-r_c|>R_b.
\end{array}\right.\,.
\end{equation}

\begin{figure}[t]
\centering
\includegraphics[width=13cm]{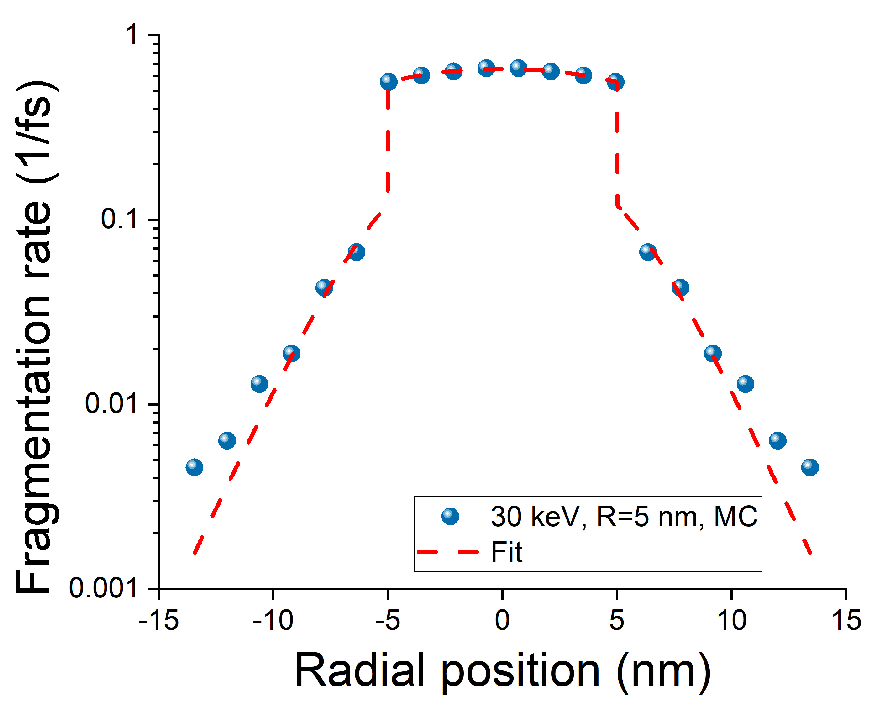}
\caption{Radial profile of the electron-induced fragmentation rate. Dots show the values obtained from Monte Carlo simulation \cite{DeVera2020}. The dashed line corresponds to the fit using Eq.~\ref{Eq:BeamProfile}. }
\label{fig:BeamProfile}
\end{figure}

\noindent
Here the constant $A$ describes the fragmentation rate caused by the primary electron beam. The constant $B$ determines the fragmentation rate caused by the secondary electrons. The quantities $r_c$ and  $R_b$ are the center of the primary electron beam and its radius respectively, and $w_0$ determines the dispersion of the secondary electron distribution. For the primary electron beam with a radius of $R_b$=5 nm and a specific electron flux of  $J_0 = $ 1 electron/(\AA$^2fs$), the best fit with  Eq. (\ref{Eq:BeamProfile}) of the fragmentation rate distribution simulated in \cite{DeVera2020} using the Monte Carlo method  was obtained with $A = 0.438$, $B = 0.241$,  $w_0  = 1.678$ nm. The comparison of the distribution from \cite{DeVera2020} with the fit derived with Eq. \ref{Eq:BeamProfile}) is shown in Fig.~\ref{fig:BeamProfile}.

The equation (\ref{Eq:BeamProfile}) with the above parameters has been used in our work to evaluate  fragmentation probabilities of the deposited W(CO)$_6$  precursors upon their irradiation with the electron beam in FEBID simulations. The results of such simulations are reported below.

In the simulations the precursor fragmentation probability was assigned to each  cell of the simulation grid according to the distribution described by Eq. (\ref{Eq:BeamProfile}). For the case study considered in our work, the grid cell size is equal to $d = 0.63$~nm, which corresponds to the size of an intact precursor molecule. This means that the considered beam diameter is covered by 16 grid cells. 
The fragmentation rates shown in Fig.~\ref{fig:BeamProfile} are obtained by dividing the values used in simulations by $3.13\times10^{-7}$.

\subsection{Protocol of atomistic MD simulations}

In the previous sections, atomistic MD simulations were used to describe the detachment process of W(CO)$_6$ and W particles from both a SiO$_2$-H substrate and a W(CO)$_6$ precursor molecule island. The protocol used for these simulations is given below.

%As explained above, two sets of atomistic MD simulations were %performed: (i) to describe the detachment process of W(CO)$_6$ and W %particles from the SiO$_2$-H substrate, and (ii) to describe the %detachment process of W(CO)$_6$ and W particles from an island of %W(CO)$_6$ precursor molecules. The following protocol was employed %for the individual MD simulations.

For the detachment of W(CO)$_6$ and W particles from the SiO$_2$-H substrate, the substrate was modelled as a 10 nm $\times$ 10 nm SiO$_2$-H patch. Simulations began with the center of the W(CO)$_6$ or W particle placed 15 {\AA} above the SiO$_2$-H surface. To improve statistical sampling, five different initial orientations of the W(CO)$_6$ molecule were considered, while the W particle was positioned over five different substrate locations. The substrate atoms were kept fixed, whereas the atoms in the W(CO)$_6$ molecules were allowed to move relative to one another. Interactions between the atoms of the SiO$_2$-H surface, W(CO)$_6$, and W particles were modelled using the CHARMM force field, which includes bonded, angular, and nonbonded interactions \cite{MacKerell1998, MacKerell2004, Huang2013,Huang2013,Beltukov2022,Verkhovtsev2021,Solovyov2012}. Table \ref{tab:SimParams} lists all the parameters employed in the simulations. W(CO)$_6$ and W particles were assigned an initial velocity of 0.007 Å/fs directed towards the substrate. MD simulations were run for 4 ps and 1.6 ps for W(CO)$_6$ and W particles, respectively, with an integration timestep of 0.1 fs and a cutoff distance of 10 {\AA} for nonbonded interactions.

\begin{table}[!t]
\caption{Interaction parameters used in atomistic MD simulations to probe the interactions of W(CO)$6$ and W particles with the SiO$2$-H substrate and with each other. Parameters follow CHARMM force field nomenclature \cite{MacKerell1998, MacKerell2004, Huang2013,Huang2013,Beltukov2022,Verkhovtsev2021,Solovyov2012} and include bonded parameters, modelled as $V(\textrm{bond})=k_b(b-b_0)$, where $b$ is the bond distance; angular parameters, modelled as $V(\textrm{angle})=k{\theta}(\theta-\theta_0)$, where $\theta$ is the bond angle; and nonbonded parameters, modelled as $V(\textrm{Lennard-Jones})=\varepsilon{ij}((r_{min_{ij}}/r_{ij})^{12}-2(r_{min_{ij}}/r_{ij})^{6})$, where $\varepsilon_{ij}=\sqrt{\varepsilon_i\varepsilon_j}$ and $r_{min_{ij}}=r_{min_{i}}/2+r_{min_{j}}/2$, with $r_{ij}$ being the distance between particles $i$ and $j$.}
\label{tab:SimParams}
\centering
\begin{tabular}{p{2cm}p{2cm}p{2cm}p{3.5cm}p{2.5cm}p{2cm} }
    \toprule
		type 1  &  type 2  &  -  &  $k_b$ (kcal/mol/\AA$^2$)   & $b_0$ (\AA) & Source\\
		\hline
		W  &  C   & -  & 127.55  & 2.11  & \cite{Cetini1963,Linstrom2001,DeVera2019}\\
		C  &  O   & -  & 1365.63 & 1.14  & \cite{DeVera2019}\\
    \hline
    \hline
		type 1  &  type 2  &  type 3  &  $k_{\theta}$ (kcal/mol/rad$^2$)   & $\theta_0$ (deg) & Source\\
    \hline
		C  &  W   &  C & 76.44   & 90.0  & adapted from \cite{Prosvetov2021}\\
        W  &  C   &  O & 14.00   & 180.0 & adapted from \cite{Prosvetov2021}\\
    \hline
    \hline
		type 1  &  -  &  -  &  $\varepsilon_i$ (kcal/mol)   & $r_{min_i}/2$ (\AA) & Source\\
    \hline
		W  &  -   &  - &  -33.475  &  1.25  & \cite{Filippova2015}\\
		C  &  -   &  - &  -0.095   &  1.95  & \cite{Mayo1990}\\        
		O  &  -   &  - &  -0.261   &  1.76  & \cite{Butenuth2012a}\\        
		Si &  -   &  - &  -0.300   &  1.60  & \cite{Butenuth2012a}\\        
		H  &  -   &  - &  -0.021   &  1.00  & \cite{Butenuth2012a}\\        
    \botrule
\end{tabular}
\end{table}

A similar protocol was followed for the detachment of W(CO)$_6$ and W particles from an island of W(CO)$_6$ precursor molecules. A cluster of 19 W(CO)$_6$ molecules with frozen internal degrees of freedom was used to model the precursor island. As above, five independent simulations were conducted for each interaction case. All other simulation parameters remained the same, except for the total simulation time, which was set to 3 ps for W(CO)$_6$ particles and 1.2 ps for W particles.

All MD simulations produced potential energy profiles as functions of the distance between the center of mass of the interacting species, similar to the examples shown in Fig.~\ref{fig:AdsorptionEnergy}. Each profile was fitted with a Morse potential around the corresponding energy minimum. Since each interaction case was simulated five times, five sets of fitting parameters were obtained. The average values of these fitting parameters were subsequently used to define the interaction parameters for the SD simulations.

\subsection{Protocol of the FEBID simulations}

The FEBID process was modeled as a sequence of 100 irradiation/replenishment cycles, each consisting of two distinct phases: (i) an irradiation phase and (ii) a replenishment phase. The duration of each irradiation phase was set to 0.31 ms, and  the duration of each replenishment phase to 0.12 ms. These correspond to a total number of simulation steps of $N_{\textrm{irrad}}=3.21\times10^6$ and $N_{\textrm{rep}} = 1.28 \times 10^6$, respectively, with the previously derived simulation time step of $\Delta t = 0.097$ ns. It is important to emphasise that the chosen durations of the simulated irradiation and replenishment cycles are consistent with those reported experimentally \cite{Fowlkes2010,Weirich2013,UTKE2022,Porrati2009}, and that all the processes involved in FEBID and discussed above are taken into account at their actual rates.

An initial injection phase was also included, during which precursor molecules were introduced into the initially empty simulation box over $N_{\textrm{inj}} = 1.0 \times 10^6$ steps. The final state of the system at the end of each phase was used as the starting configuration for the subsequent phase.

Simulations were performed on a cubic lattice initially consisting of $51 \times 51 \times 21$ unit cells. As the simulation progressed and material was deposited, the simulation box was dynamically extended in the z-direction. This was done to ensure that the free volume for adding new precursor particles remained constant throughout the FEBID process. After each irradiation phase, the local deposit height was computed based on the distribution of adsorbed and immobilized species, and the simulation box height was increased accordingly. This adaptive approach prevented artificial confinement effects and maintained physically realistic conditions for precursor injection at each cycle, mimicking the continuous exposure environment encountered in experimental FEBID setups.

The physical processes underlying FEBID, as described in the preceding sections, were incorporated into the simulations. The probabilities associated with each process are summarized in the accompanying tables above and were used as input parameters in the SD simulations performed using MBN Explorer \cite{Solovyov2012,Solovyov2022}. In particular, the irradiation phase included particle motion, deposition and fragmentation, while the replenishment phase excluded the possibility of particle fragmentation.

%%%%%%%%%%%%%%%%%%%%%%%%%%%%%%%%%%%%%%%%%%%%%%%%%%%%%%%%%%%%%%%%%%%
%%%%%%%%%%%%%%%%%%%%%%%%%%%%%%%%%%%%%%%%%%%%%%%%%%%%%%%%%%%%%%%%%%%
\section{Results and discussion}
\label{Results}

\begin{figure}[htp!]
\centering
\includegraphics[width=13cm]{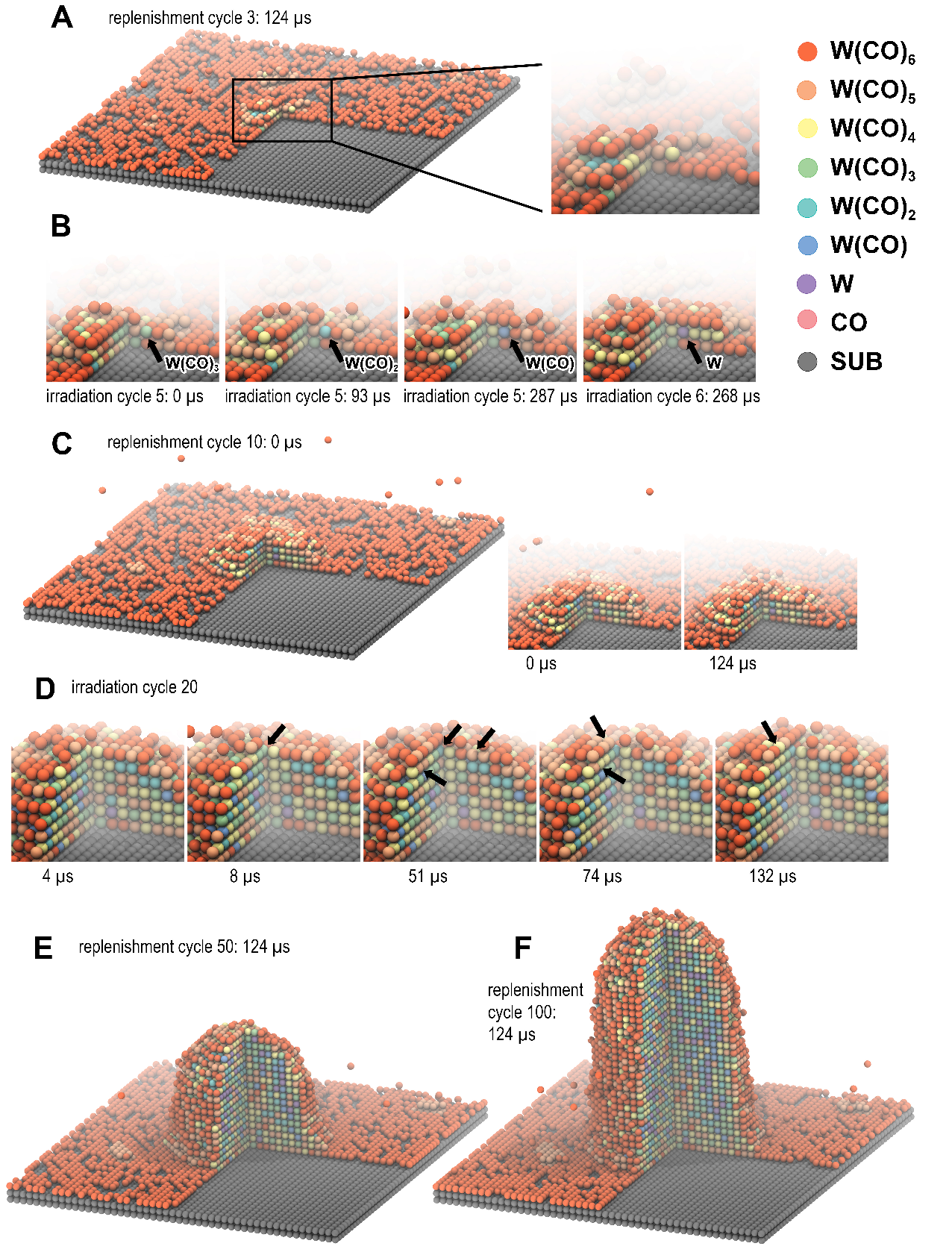}
\caption{Evolution of a W-rich structure emerging on a SiO$_2$ surface after 3 irradiation/replenishment FEBID cycles (\textbf{A}), during the 5th and the 6th irradiation cycle (\textbf{B}), at the beginning and at the end of the 10th replenishment cycle (\textbf{C}), during the 20th irradiation cycle (\textbf{D}), after the 50th replenishment cycle (\textbf{E}), and after the 100th replenishment cycle (\textbf{F}). The time instances that correspond to the rendered snapshots are indicated. Arrows in panels \textbf{B} and \textbf{D} indicate particles that change types in the course of the irradiation. Colour indicates particle types and was chosen consistent with the ones defined in Fig.~\ref{fig:introParticles}. The entire process of structural evolution of the formed nanoparticle is also shown in Supplementary video S1.} 
\label{fig:Snapshot}
\end{figure}

Figure~\ref{fig:Snapshot} presents a sequence of three-dimensional visualizations of the W-rich nanostructure evolution on a SiO$_2$-H substrate, modeled via SD simulations of the FEBID process. The panels illustrate the spatial and compositional development of the deposit across selected irradiation/replenishment cycles, providing insight into the dynamic interplay of surface reactions, particle motion, and electron-induced precursor dissociation under realistic FEBID conditions. The Supplementary video 1 furthermore illustrates the growing process dynamically.

Figure~\ref{fig:Snapshot}A shows the system after 3 full FEBID cycles, revealing the initial adsorption and fragmentation of precursor molecules within the region irradiated by the focused electron beam. The early-stage deposit is characterised by the formation of a thin, disordered layer made up of precursors and their fragments. A few isolated clusters have formed at the spot where the beam hits the substrate. At this stage, surface coverage is sparse, and the vertical growth is minimal.

Figures~\ref{fig:Snapshot}B-C track the deposit during the 5th to 10th cycles, highlighting the key features of the process at this stage. Figure~\ref{fig:Snapshot}B captures the 5th and the 6th irradiation cycles, where particle fragmentation is clearly seen. Arrows indicate individual selected precursor molecules undergoing dissociation, resulting in the release of volatile CO ligands and retention of W-rich fragments on the substrate. Figure~\ref{fig:Snapshot}C illustrates the start and the end of the 10th replenishment phase. A comparison of Figs.~\ref{fig:Snapshot}A and C shows the appearance of the three-dimensional structure of the deposit.  
It can also be seen that all the precursors above the surface have disappeared by the end of this period.

Figure~\ref{fig:Snapshot}D, representing the 20th irradiation cycle, demonstrates the growing complexity of the nanostructure. The accumulation of metal-containing species is now concentrated near the center of the electron beam spot, with increasing density and vertical build-up. The formation of spatially structured features becomes apparent, reflecting the interplay of beam-induced reactions and constrained diffusion pathways.

Figure~\ref{fig:Snapshot}E shows the deposit after 50 cycles. At this intermediate stage, a well-defined bell-shaped structure is formed. The central region displays dense accumulation of tungsten-rich fragments, surrounded by a less dense periphery formed from partially fragmented precursors, undergoing the lateral diffusion.
This morphology is consistent with the expected distribution of primary and secondary electron flux, where the central zone experiences the highest fragmentation rates.

Finally, Figure~\ref{fig:Snapshot}F depicts the state of the system following 100 complete FEBID cycles. At this point, a pronounced nanostructure with a peak height of approximately 15 nm has emerged. The structure exhibits a clear gradient in particle composition and spatial distribution, with the core composed of highly fragmented W(CO)$_6$ derivatives, while the peripheral regions show evidence of metal-rich islands and dispersed fragments. The appearance of these peripheral features is attributed to the action of secondary electrons and the stochastic diffusion of particle fragments beyond the primary beam spot.

The visualizations use a consistent colour scheme defined in Figure~\ref{fig:introParticles}, where each particle type—intact precursors, fragmented intermediates (W(CO)$_{6-1}$), pure tungsten atoms, CO ligands, and substrate particles are represented distinctly. The SD framework captures key kinetic events such as particle injection, diffusion, dissociation, and detachment, all governed by rate constants derived from atomistic molecular dynamics simulations and experimental parameters, as described in the previous section.

Figure~\ref{fig:Snapshot} shows that the SD approach to multiscale FEBID modelling produces a realistic temporal evolution of a nanoscale structure, including the vertical and lateral growth trends, particle type distributions, and emergent morphologies. These simulation results are qualitatively consistent with experimental observations \cite{Fowlkes2010,Weirich2013,UTKE2022,Porrati2009} and emphasise the importance of capturing both primary beam effects and long-range secondary electron interactions in predictive FEBID simulations. The results shown in Fig.~\ref{fig:Snapshot} were obtained using parameter Set 3 from Table~\ref{tab:my_labelPREC}.

%\textcolor{red}{Average metal content and height of the deposit. Correspondence ot the beam profile. }

\begin{figure}[t!]
\centering
\includegraphics[width=15cm]{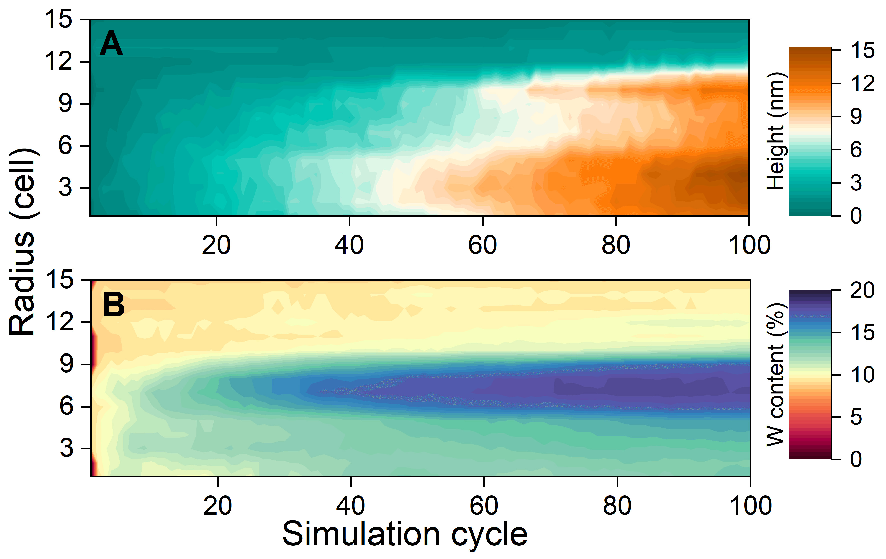}
\caption{The figure shows how  the average height (\textbf{A}) and metal content (\textbf{B}) of the nanostructure evolve and become distributed as a function of radius  following an increasing number of irradiation/replenishment FEBID cycles.
Colour indicates either the deposit height (\textbf{A}) or the metal content (\textbf{B}) calculated as average values over grid cells located at a fixed distance from the electron beam center. The values presented are the time-averaged heights and metal contents over each completed irradiation/replenishment cycle. }
\label{fig:Growth_Evolution}
\end{figure}

The nanostructure shown in Fig.~\ref{fig:Snapshot} is relatively small compared to the sizes of structures that can be fabricated by FEBID \cite{Fowlkes2010,Weirich2013,UTKE2022}. However, it is large enough to determine the two essential characteristics of the FEBID process:  the growth rate (i.e the height of the grown nanostructure as a function of time or the number of FEBID cycles), and the metal content of the nanostructure. These characteristics were evaluated for the three simulated scenarios, see Table~\ref{tab:my_labelPREC}. Figure~\ref{fig:Growth_Evolution} shows the growth evolution of the calculated radial profiles for the deposit height and metal content in the case of parameter Set 3.  The deposit's height and metal content were analysed radially, under the assumption that the electron beam hits the substrate at its center.

%The nanostructure shown in Fig.~\ref{fig:Snapshot} corresponds to an elementary nanostructure grown in the course of the FEBID process, which could be considered as a ``pixel'' for the growth of a more complex structure \cite{...}. The two main characteristics of this elementary nanostructure are the height of the deposit and its metal content \cite{...}. Both characteristics were evaluated for the three simulated scenario, see Table~\ref{tab:my_labelPREC}. Figure~\ref{fig:Growth_Evolution} shows the growth-evolution of the calculated radial profiles for the deposit height and metal content in the case of parameter Set 3. The analysis assumed that the electron beam hits the substrate at its center, and the deposit's height and metal content were analysed radially.

Note that the size of the grid cells was fixed in the simulations, which creates a slight discrepancy between the volume of one grid cell and the actual physical volume of particles placed therein. To mitigate this discrepancy in the analysis of the FEBID process, the volume of each simulated particle was evaluated as a sum of volumes of its constituent atoms calculated from the van der Waals radii taken as R$_{\rm vdW}$(W) = 0.21 nm, R$_{\rm vdW}$(C) = 0.17 nm, R$_{\rm vdW}$(O) = 0.152 nm \cite{PubChem}.

The characteristic size $d$ of one grid cell was set to the size of the largest particle in the system, the precursor molecule W(CO)$_6$. The value of $d$ was estimated as $d=V_{W(CO)6}^{1/3}$, with $V_{W(CO)6}$ being the volume of W(CO)$_6$. The necessary analysis also included the projection area of a particle on the substrate plane ($x$, $y$), which was assumed to be the same for all particles in the system, being taken as the projection area of the W(CO)$_6$ particle, calculated as $S=d^2$. The values relevant to the present investigation are summarized in Table~\ref{tab:ParticleSize}. The total height of the deposit at each location of the substrate was calculated from the individual height contributions of the individual particles by adding the heights at each point ($x$, $y$) from the substrate to the deposit surface.

\begin{table}[h!]
\caption{The characteristic volumes $V$ and the average heights $h$ of the different particles used to model the FEBID process. The projection area $S = 0.397$~nm$^2$ was assumed to be the same for precursor particles and their derivatives.}
	\label{tab:ParticleSize}
	\centering
	\begin{tabular}{p{2cm}p{1.5cm}p{1.5cm}p{1.5cm}}
		\toprule
		Particle    &   V (nm$ ^3 $) & $h$ (nm) \\
		\hline
		W(CO)$_6$	& 0.251 & 0.630	\\   %	0.630 	& 0.397  \\
		W(CO)$_5$	& 0.215 & 0.541	\\   %	0.599   & 0.359  \\
		W(CO)$_4$	& 0.180 & 0.453	\\   %  0.565   & 0.319  \\
		W(CO)$_3$	& 0.145 & 0.365	\\   %	0.525   & 0.276  \\
		W(CO)$_2$	& 0.109 & 0.274	\\   %	0.478   & 0.229  \\
		W(CO)	    & 0.074 & 0.186	\\   %  0.420   & 0.176 \\
		W		    & 0.039 & 0.098	\\   %  0.339  	& 0.115  \\
		CO          & 0.035 & 0.088  \\   %  0.328   & 0.108  \\		
	\botrule
	\end{tabular}
\end{table}

The height of the deposit in Fig.~\ref{fig:Growth_Evolution} was calculated for each irradiation/replenishment cycle as the radial average. The percentage of the metal content was calculated as the mass fraction of tungsten atoms relatively to the other elements (C, O) contained in the simulation box cells placed radially from the center of the electron beam. Here the volumes of the individual atoms were assumed based on the van der Waals radii, given in Table~\ref{tab:ParticleSize}. Figure~\ref{fig:Growth_Evolution}A shows a color-coded spatial map of the deposit height across the substrate. As the number of FEBID cycles increases, the height profile exhibits a progressive and localized growth at the beam center. This results in the formation of a dome-shaped structure, peaking directly under the beam and tapering off smoothly with radial distance. The deposit height increases nonlinearly over time, reflecting the interplay of precursor adsorption, electron-induced dissociation, and limited diffusion. After 100 cycles, the maximum deposit height reaches approximately 15 nm, consistent with the expected vertical growth in experiments using similar beam parameters \cite{Fowlkes2010,Porrati2009,Huth2012}.

Figure~\ref{fig:Growth_Evolution}B shows the radial profile of the metal content. Interestingly, the metal content does not peak at the beam center but instead reaches its maximum at an intermediate radial distance—forming a distinct ring-like feature indicative of a ``crown-shaped'' composition profile. This spatial inhomogeneity arises from the constrained mobility of ligands and fragment particles within the central region of the growing deposit. As the structure thickens at the center, the dense network of particles increasingly restricts the escape of CO ligands, suppressing further dissociation and reducing the net metal fraction. In contrast, regions at the beam periphery remain more accessible, allowing for efficient ligand escape and higher metal accumulation.

The simulation results reflect the underlying stochastic dynamics model, where fragmentation events depend not only on electron flux but also on local spatial availability for dissociated ligands to occupy neighboring cells. The crown-like metal distribution is in qualitative agreement with prior experimental observations of metal-enriched rings surrounding FEBID structures, often attributed to secondary electron effects and transport-limited fragmentation dynamics \cite{Huth2012,vanDorp2008}.

\begin{figure}[t!]
\centering
\includegraphics[width=15cm]{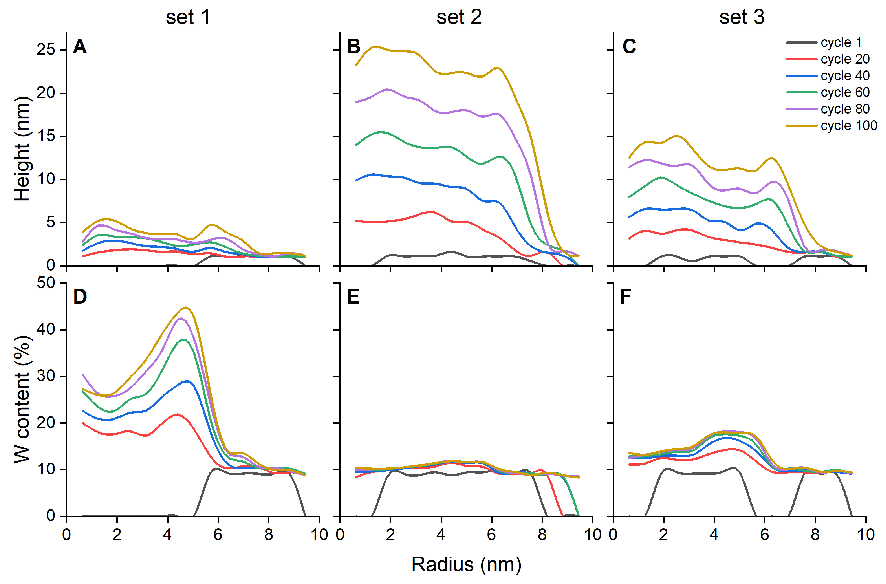}
\caption{Radial distributions of the height of the deposit (\textbf{A}, \textbf{B}, \textbf{C}) and the corresponding metal content (\textbf{D}, \textbf{E}, \textbf{F}) calculated for the deposits after 1 (gray line), 20 (red line), 40 (blue line), 60 (green line), 80 (purple line), and 100 (ochre line) irradiation/replenishment cycles. Results are shown for the simulations performed with parameter sets 1-3, see Table~\ref{tab:my_labelPREC}.}
\label{fig:Radial_Height_Content}
\end{figure}

It is also interesting to analyze the evolution of the deposit height and its metal content at specific time instances of the simulated FEBID process. Here it is particularly possible to compare the results of the simulations obtained for the different sets of the parameters, see Table~\ref{tab:my_labelPREC}. Figure~\ref{fig:Radial_Height_Content} shows the radial distributions of the height of the deposit and the corresponding metal content for the three simulation parameter sets.

Figure~\ref{fig:Radial_Height_Content}A–C display the temporal evolution of the radial height profiles of the deposits. For all parameter sets, the growth initiates at the beam center and expands radially and vertically as the number of FEBID cycles increases, ultimately forming a bell-shaped profile. The growth rate and final structure dimensions, however, differ significantly among the three sets. Set 1 leads to slower and more confined deposition, while Set 2 yields the fastest growth rate and the tallest structures. Simulation with Set 3 provide an intermediate growth regime with a nanostructure, reaching approximately 15 nm after 100 cycles—comparable to experimental data for similar conditions \cite{Fowlkes2010,Porrati2009,Huth2012}.

Figure~\ref{fig:Radial_Height_Content}D–F show the corresponding radial distributions of the metal content. In all cases, the W-content peaks not at the beam center but in a surrounding annular region, forming a ``crown-like'' profile. This feature, however is minimally visible in the case of parameter Set 2.

Among the three parameter sets, Set 3 stands out as the most physically realistic and experimentally consistent. It yields a final tungsten content in the central deposit region of approximately 15 \% and a total deposit height in the 15 nm range after 100 cycles—values that align well with known experimental measurements of W(CO)$_6$-based FEBID deposits. Sets 1 and 2 either under/overestimate the metal content or lead to significantly reduced/enhanced growth rates, suggesting that they do not adequately capture the balance between precursor fragmentation, ligand detachment, and molecular diffusion effects within the growing nanostructure.

Fig.~\ref{fig:Radial_Height_Content} shows detailed insights into the spatial and temporal evolution of the FEBID deposit.  The  analysis presented confirms that parameter set 3 is  the most accurate model for simulating realistic nanostructure formation.  The consistency   of the simulated  geometries and compositions with those observed in the experiment further reinforces the reliability of the underlying SD framework and its applicability to predictive FEBID modelling.

\begin{figure}[t!]
\centering
\includegraphics[width=0.6\textwidth]{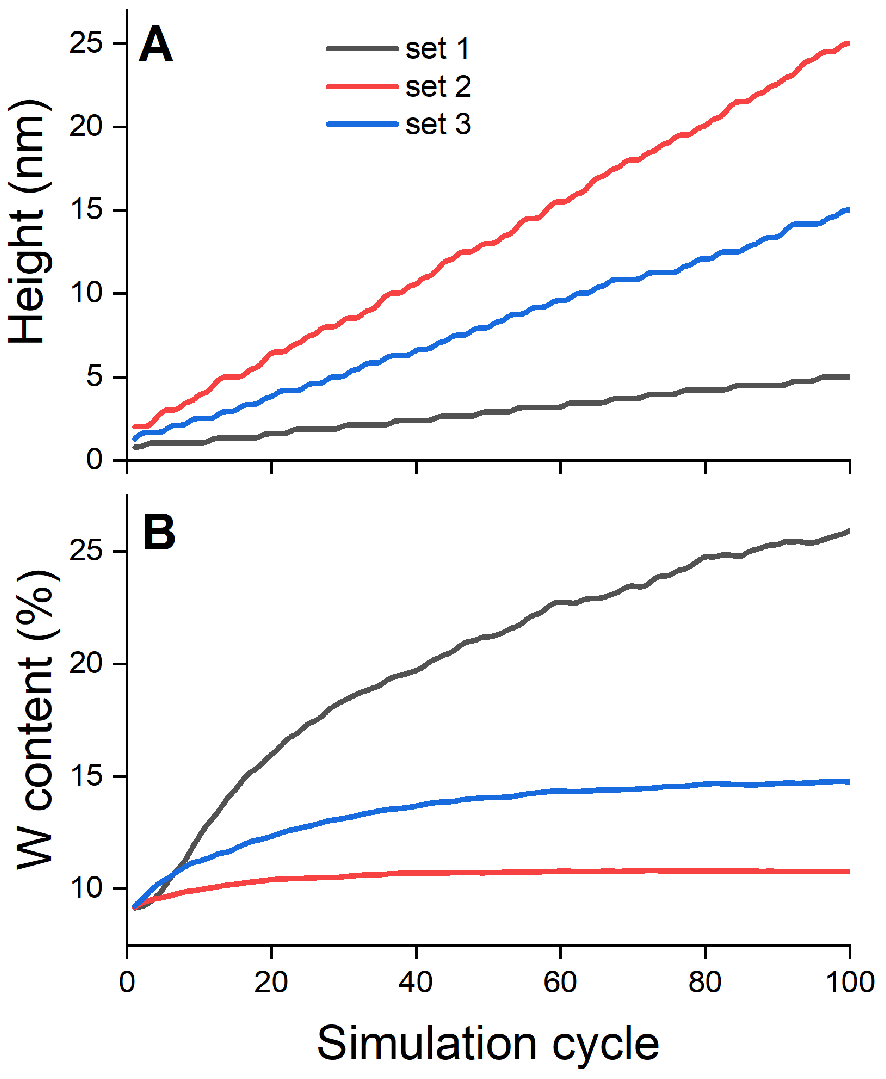}
\caption{Evolution of the height of the deposit (\textbf{A}) and the average metal content (\textbf{B}) and as a function of the number of the simulated FEBID cycles. Results are shown for the simulations performed with parameter sets 1-3, see Table~\ref{tab:my_labelPREC}.}
\label{fig:Growth_Evolution_2}
\end{figure}

Figure \ref{fig:Growth_Evolution_2} displays the temporal evolution of the average height of the deposit (panel A) and the corresponding average tungsten content (panel B), both evaluated as functions of the number of irradiation/replenishment cycles. The results are shown for all three parameter sets used in the SD simulations, with particular attention to parameter Set 3, which yields the most experimentally realistic behaviour \cite{Fowlkes2010,Porrati2009,Huth2012}.

Figure \ref{fig:Growth_Evolution_2}A shows the steady increase in deposit height over the course of 100 FEBID cycles. For parameter set 3, the deposit reaches a height of approximately 15 nm after 100 cycles. Given that each cycle consists of an irradiation phase (0.31 ms) and a replenishment phase (0.12 ms), the total simulated time for 100 cycles is 43 ms. This value allows to evaluate the vertical growth rate of the nanostructure for parameter Set 3 as 0.35 nm/ms.

This value is consistent with  with experimentally reported growth rates for FEBID using W(CO)$_6$ precursors and similar electron beam parameters, where vertical deposition rates typically range from 200 to 500 nm/s, depending on beam current, pressure, and surface conditions \cite{Fowlkes2010,Weirich2013,UTKE2022,Porrati2009,Huth2012}.

Figure \ref{fig:Growth_Evolution_2}B depicts the evolution of the average tungsten content in the deposit over time. Parameter Set 3 again yields the most realistic outcome, with the W content stabilizing in the range of $\sim$15 \% after an initial transient phase. This value is consistent with experimental measurements of tungsten concentrations in FEBID structures grown from W(CO)$_6$, where typical metal contents fall within 25–40\%, depending on deposition conditions and beam-induced fragmentation efficiency \cite{Fowlkes2010,Weirich2013,UTKE2022,Porrati2009,Huth2012}.

%%%%%%%%%%%%%%%%%%%%%%%%%%%%%%%%%%%%%%%%%%%%%%%%%%%%%%%%%%%
%%%%%%%%%%%%%%%%%%%%%%%%%%%%%%%%%%%%%%%%%%%%%%%%%%%%%%%%%%%
\section{Conclusions}
\label{Conclusions}

We have reported on  a new approach to the multiscale computational modelling of the focused electron beam-induced deposition (FEBID), realised using the advanced software packages: MBN Explorer and MBN Studio. Our approach is based on stochastic dynamics (SD), which describes the probabilistic evolution of complex systems. The parameters for the new SD-based FEBID model were determined using molecular dynamics (MD) simulations. We have developed a new methodology for this purpose, which  is described in detail. This methodology can be applied to many other case studies of the dynamics of complex systems. Our work focuses on the FEBID process involving W(CO)$_6$ precursor molecules deposited on a hydroxylated SiO$_2$-H substrate.  Simulations and a detailed analysis of a growing W-rich nanostructure were performed. This new approach was shown to provide essential atomistic insights into the complex FEBID process, including the elemental composition and morphology of the deposit at each stage of growth. The derived results were then compared with experimental observations and validated.

The multiscale approach and related methodologies developed in this study can be applied to many other FEBID case studies. These include a variety of different precursor molecules, electron beam parameters and  substrates used in FEBID. In this way, the FEBID chemistry, which sometimes involves also additional molecules such as H$_2$O, can be studied in great detail. The SD method can also be used to model  a variety of many other growth processes and  fabrication techniques for nanostructures, such as atomic layer deposition, chemical and physical vapor deposition. These techniques can be exploited with or without radiation and involve plasma. The SD methodology has many applications in modelling biological systems, chemistry and materials science.  It can be used for multiscale computational modelling of a large number of case studies; see the roadmap \cite{Solovyov2024} and references therein.

The method can also be used to model  a variety of many other growth processes and  fabrication techniques for nanostructures, such as ALD, CVD and PVD, which can be exploited with or without radiation or involving plasma. It has many applications in modelling biological systems.

The multiscale methods developed in this study can be  further upgraded and applied to important technological developments, such as 3D nanoprinting.

%%%%%%%%%%%%%%%%%%%%%%%%%%%%%%%%%%%%%%%%%%%%%%%%%%%%%%%%%%%
%%%%%%%%%%%%%%%%%%%%%%%%%%%%%%%%%%%%%%%%%%%%%%%%%%%%%%%%%%%
\section{Supporting Information Available}
\label{SupInfo}

\noindent
\textbf{Supporting Video S1:} illustrates the dynamic process of structural evolution of the W-rich nanoparticle atop the SiO$_2$-H surface. The characteristic temporal and spatial scales of the underlying processes are indicated. A sector of the emerging nanoparticle is not shown to better illustrate the internal processes occurring inside. The green transparent cylinder illustrates the electron beam that is present during the irradiation phases.

\section{Data Availability Statement}
\label{DataAvail}
Data underpinning this publication are available from the Supplementary material. MD and SD trajectories will be provided upon reasonable request.

%%%%%%%%%%%%%%%%%%%%%%%%%%%%%%%%%%%%%%%%%%%%%%%%%%%%%%%%%%%%%%%%%%%%%%%%%%%%%%%%%%%%%%%%

\begin{acknowledgments}
The authors would like to thank the Volkswagen Foundation (Lichtenberg professorship awarded to I.A.S.), the Deutsche Forschungsgemeinschaft (SFB 1372 Magnetoreception and Navigation in Vertebrates, no. 395940726 to I.A.S.; TRR386 HYP*MOL, no. 514664767 to I.A.S.), the Ministry of Science and Culture of Lower Saxony (Simulations Meet Experiments on the Nanoscale: Opening up the Quantum World to Artificial Intelligence (SMART) to I.A.S; and Dynamik auf der Nanoskala: Von kohärenten Elementarprozessen zur Funktionalität (DyNano) to I.A.S). We would like to express our gratitude to the European Commission for its partial financial support of this work through the H2020-MSCA-RISE-2019 RADON project (grant agreement no. 872494).
This article is also based on work conducted as part of COST Action CA20129 ``Multiscale Irradiation and Chemistry Driven Processes and Related Technologies'' (MultIChem), supported by COST (European Cooperation in Science and Technology). 

\end{acknowledgments}

\bibliographystyle{nature}
\bibliography{MBN-RC,MBN-RC_add}

\end{document}